\documentclass[aps,nofootinbib,notitlepage,longbibliography,twocolumn, superscriptaddress,preprintnumber]{revtex4-1}
\usepackage{amsmath,amssymb,amsfonts}
\usepackage{slashed}
\usepackage{graphicx}
\usepackage{bm}
\usepackage{float}
\usepackage[T1]{fontenc}
\usepackage{multirow}
\usepackage[utf8]{inputenc}
\usepackage{gauss} 
\usepackage[normalem]{ulem}
\usepackage[top=1.0in,bottom=1.0in,left=1.0in,right=1.0in]{geometry}
\usepackage[colorlinks,linkcolor=blue,citecolor=blue,urlcolor=blue]{hyperref}
\usepackage{graphicx}
\usepackage{subcaption}
\usepackage{ragged2e}
\usepackage[compatibility=false]{caption} 
\DeclareCaptionJustification{justified}{\justifying}
\usepackage{subcaption}
\usepackage{booktabs}
\usepackage{caption}
\usepackage{qcircuit}
\newcommand{\be}{\begin{equation}}
\newcommand{\ee}{\end{equation}}
\newcommand{\bea}{\begin{eqnarray}}
\newcommand{\eea}{\end{eqnarray}}
\newcommand{\ba}{\begin{eqnarray}}
\newcommand{\ea}{\end{eqnarray}}

\newcommand{\ket}[1]{\left|#1\right\rangle}
\newcommand{\bra}[1]{\left\langle#1\right|}

\def\be{\begin{eqnarray}}
\def\ee{\end{eqnarray}}
\def\bea{\be}
\def\eea{\ee}

\def\roughly#1{\mathrel{\raise.3ex\hbox{$#1$\kern-.75em%
\lower1ex\hbox{$\sim$}}}}

\def\abs#1{{\left| #1 \right|}}

\usepackage{lipsum}

\usepackage{fancyhdr}

\preprint{NT@UW-25-17}
\begin{document}

\title{Tensor network simulations of quasi-GPDs in the massive Schwinger model}
\author{Sebastian Grieninger}
\email{segrie@uw.edu}
\affiliation{InQubator for Quantum Simulation (IQuS), Department of Physics, University of Washington, Seattle, WA 98195}
\affiliation{Department of Physics, University of Washington, Seattle, WA 98195}
\affiliation{Co-design Center for Quantum Advantage (C2QA), Stony Brook University, Stony Brook, New York 11794–3800, USA}
\affiliation{Center for Nuclear Theory, Department of Physics and Astronomy, Stony Brook University, Stony Brook, New York 11794–3800, USA}

\author{Jake Montgomery}
\email{jake.montgomery@stonybrook.edu}
 \affiliation{Center for Nuclear Theory, Department of Physics and Astronomy,
Stony Brook University, Stony Brook, New York 11794–3800, USA}
\author{Felix Ringer}%
 \email{felix.ringer@stonybrook.edu}
\affiliation{Center for Nuclear Theory, Department of Physics and Astronomy, Stony Brook University, Stony Brook, New York 11794–3800, USA}
\author{Ismail Zahed}
\email[]{ismail.zahed@stonybrook.edu}
\affiliation{Center for Nuclear Theory, Department of Physics and Astronomy,
Stony Brook University, Stony Brook, New York 11794–3800, USA}

\begin{abstract}
Generalized Parton Distribution functions (GPDs) are off-diagonal light-cone matrix elements that encode the internal structure of hadrons in terms of quark and gluon degrees of freedom. In this work, we present the first nonperturbative study of quasi-GPDs in the massive Schwinger model, quantum electrodynamics in 1+1 dimensions (QED2), within the Hamiltonian formulation of lattice field theory. Quasi-distributions are spatial correlation functions of boosted states, which approach the relevant light-cone distributions in the luminal limit. Using tensor networks, we prepare the first excited state in the strongly coupled regime and boost it to close to the light-cone on lattices of up to 400 lattice sites. We compute both quasi-parton distribution functions and, for the first time, quasi-GPDs, and study their convergence for increasingly boosted states. In addition, we perform analytic calculations of GPDs in the two-particle Fock-space approximation and in the Reggeized limit, providing qualitative benchmarks for the tensor network results. Our analysis establishes computational benchmarks for accessing partonic observables in low-dimensional gauge theories, offering a starting point for future extensions to higher dimensions, non-Abelian theories, and quantum simulations.
\end{abstract}

\maketitle
\tableofcontents 
\thispagestyle{fancy}
\fancyhead{}
\fancyhead[R]{IQuS@UW-21-115, NT@UW-25-17}
\fancyfoot{}
\renewcommand{\headrulewidth}{0pt}

\section{Introduction}
\label{SEC1}
The internal structure of protons, neutrons, and heavier nuclei is encoded in light-cone distributions, which play a central role in describing high-energy scattering processes in Quantum Chromodynamics (QCD). These functions characterize how the momentum and spin of a hadron are distributed among its elementary quark and gluon constituents, which can be studied at high-energy collider experiments. When QCD factorization holds, the cross section of a hard scattering process can be expressed as a perturbatively calculable coefficient function convolved with the relevant parton distributions or fragmentation functions, as in the case of Drell–Yan and jet production~\cite{Collins:1989gx}. Precise determinations of light-cone distribution functions are therefore critical not only for advancing our understanding of QCD, but also for connecting theory to measurements at experimental facilities. Current efforts include analyses of semi-inclusive processes at Jefferson Lab and, in the near future, at the Electron–Ion Collider (EIC)~\cite{AbdulKhalek:2021gbh,Abir:2023fpo}.

Parton distribution functions (PDFs) are defined on the light front and are intrinsically non-perturbative, which makes them inaccessible to traditional Euclidean lattice formulations beyond their lowest moments. This limitation was overcome by Ji’s proposal in Ref.~\cite{Ji:2013dva}, which introduced quasi-parton distributions. In this framework, the leading-twist light-front correlators are replaced by equal-time correlators evaluated for hadronic states boosted to large momenta. These quasi-distributions can then be matched perturbatively to their light-cone counterparts~\cite{Zhang:2017bzy,Ji:2015qla,Bali:2018spj,Alexandrou:2018eet,Izubuchi:2019lyk,Izubuchi:2018srq}. 
	
Quasi-parton distribution matrix elements computed on a finite Euclidean lattice have been shown to match, to all orders in perturbation theory, the results obtained through LSZ reduction in continuum Minkowski QCD~\cite{Briceno:2017cpo}. Variants of this approach have also been formulated, including pseudo-distributions~\cite{Radyushkin:2017gjd,Orginos:2017kos} and lattice cross sections~\cite{Ma:2014jla}. In recent years, several lattice QCD collaborations have implemented these ideas and succeeded in numerically extracting light-cone parton distributions. Furthermore, two-dimensional model studies provide nonperturbative support for the approach~\cite{Ji:2018waw}.

The present work extends our previous investigations of partonic structure in QED$_2$ using the Hamiltonian formulation and exact diagonalization techniques. In particular, in  Ref.~\cite{Grieninger:2024cdl}, we computed the quasi-parton distribution functions (qPDFs) for the lightest $\eta$' meson, examining the transition from strong to weak coupling and comparing boosted spatial distributions to light-front results. In Ref.~\cite{Grieninger:2024axp}, we introduced and computed quasi-fragmentation functions using an equal-time, spatially boosted formulation. Building on these studies, we now turn to Generalized Parton Distributions (GPDs). 

GPDs provide more detailed information about the partonic structure of hadrons. They are central to the 3D hadron tomography program that is being developed at the future EIC. In contrast to PDFs, which only describe the longitudinal momentum distribution of quarks and gluons, GPDs capture the correlations between the longitudinal parton momentum and their transverse spatial position. They provide for a three-dimensional image of hadron structure. 

Formally, GPDs are defined as off-diagonal matrix elements of leading-twist operators, and can be studied for both unpolarized and polarized hadronic targets.  We note that in 1+1 dimensions, there is no transverse dimension. However, the 1+1-dimensional version of GPDs can still be defined in terms of the longitudinal momentum fraction and the skewness induced by the momentum transfer. 2D GPDs interpolate between the partonic description
(small skewness) and the form factor description (large skewness) of two-dimensional hadrons.

Quasi-GPDs (qGPDs) are currently being investigated through lattice QCD simulations~\cite{Lin:2020rxa,Lin:2021brq,Bhattacharya:2022aob,Bhattacharya:2023ays,Bhattacharya:2023jsc,Holligan:2023jqh}, phenomenological modeling~\cite{Ji:1997gm,Boffi:2002yy,Scopetta:2003et,Petrov:1998kf,Guidal:2004nd,Radyushkin:1998bz}, and dual gravity approaches~\cite{Mamo:2022jhp,Mamo:2024jwp,Mamo:2024vjh,Hechenberger:2025rye,Hechenberger:2025wnz,deTeramond:2018ecg,Mondal:2015uha}, each contributing complementary insights into their nonperturbative structure.  For completeness, we note the recent analysis of GPDs in the
context of two-dimensional QCD in the large-$N_c$ limit~\cite{Jia:2024atq}, following on the early
work in~\cite{Burkardt:2000uu}.

In this work, we present the first non-perturbative study of qGPDs in the massive Schwinger model, quantum electrodynamics in 1+1 dimensions (QED2), using the Hamiltonian formulation of lattice field theory~\cite{Kogut:1974ag,Susskind:1976jm}.
Space is discretized on a lattice, while time remains a continuous variable. In 1+1 dimensions with open boundary conditions, the U(1) gauge field can be eliminated via Gauss’s law, and the fermion fields can be mapped to spin degrees of freedom through a Jordan–Wigner transformation~\cite{Jordan:1928wi}. This yields a spin-lattice Hamiltonian with asymmetric long-range interactions~\cite{Martinez:2016yna,Banerjee:2012pg,Klco:2018kyo}. 

Our goals are to quantify the practical computational cost of simulating nonperturbative hadronic correlation functions, explore extrapolations to the continuum and infinite volume, establish benchmark results for future quantum simulations, and provide a starting point for extensions to higher-dimensional and non-Abelian gauge theories. The analysis presented here is in line with and extends recent model studies of PDFs in 1+1 dimensions~\cite{Li:2021kcs,Li:2022lyt,Grieninger:2023ufa,Grieninger:2024cdl,Banuls:2025wiq,Kang:2025xpz}. In the context of quantum simulations, partonic observables were considered in~\cite{Echevarria:2020wct, Lamm:2019uyc,Perez-Salinas:2020nem,Briceno:2020rar,Chen:2025zeh,Galvez-Viruet:2025rmy,Galvez-Viruet:2025ket}. Simulating quantum field theories in the light-front formulation was explored in~\cite{Kreshchuk:2020dla}. 

Ultimately, quantum simulations are expected to overcome the limitations of classical methods, particularly for problems involving real-time dynamics in fundamental nuclear and particle physics~\cite{Beck:2023xhh}. A critical step in this direction was presented in Ref.~\cite{Jordan:2012xnu}, where a polynomially scaling sampling algorithm was presented for scattering processes in $\phi^4$ scalar field theory, a result expected to be beyond the reach of classical computations. Instead of targeting calculations of the full scattering process, in this work, we focus on low-energy, nonperturbative quantities such as PDFs and GPDs that can be isolated from perturbative dynamics at high energies within QCD factorization. While this approach does not provide a universal description of the full scattering process, it identifies key nonperturbative functions relevant to experiments, offering a potentially more near-term avenue toward demonstrations of quantum utility in fundamental physics. 

Here, we perform classical computations using Matrix Product States (MPS)~\cite{1992CMaPh.144..443F,PhysRevB.55.2164,PhysRevLett.75.3537}. This approach is particularly well-suited for gapped Hamiltonians in 1+1 dimensions. Ground and excited states can be prepared efficiently using the Density Matrix Renormalization Group (DMRG)~\cite{PhysRevLett.69.2863,RevModPhys.77.259} algorithm, and unitary evolution can be performed with the Time-Dependent Variational Principle (TDVP)~\cite{Haegeman:2011zz}. Previous studies have employed tensor networks and quantum simulations to investigate ground-state properties and real-time dynamics of wave packets in the Schwinger model~\cite{PhysRevResearch.4.043133,PhysRevLett.113.091601,deJong:2021wsd,Banuls:2013jaa,Farrell:2023fgd,Farrell:2024fit, Thompson:2021eze,Barata:2023jgd,Gustafson:2024bww, Florio:2025hoc,Shao:2025obi,Shao:2025ygy,Artiaco:2025qqq,Grieninger:2025rdi}, and the emergence of an out-of-equilibrium fluid phase in the transition from strong to weak coupling~\cite{Janik:2025bbz}.

The purpose of this work is to focus on PDFs and GPDs, which are defined in terms of single-particle states with definite momentum. Specifically, we prepare the first excited state of the Schwinger model in the strongly coupled regime and boost it to finite momentum. The properties of this boosted state are benchmarked against the relativistic dispersion relation, where we find good agreement with the expected result in the continuum. With MPS, we reach system sizes of up to 400 staggered lattice sites (corresponding to 200 physical sites), enabling detailed studies of quasi-PDFs and, for the first time, quasi-GPDs. We analyze their convergence with the boost parameter as the correlation functions approach the light-cone limit. In addition, we complement these results with analytic calculations of GPDs in the two-particle Fock-space approximation, which provide qualitative insight into the tensor network computations.

The outline of the paper is as follows: In section~\ref{SEC2}, we briefly review the
general features of the strong and weak coupling regime of the massive Schwinger model. In section~\ref{SEC3}, we first recall the light-front equation for the lowest meson state
in the lowest Fock state approximation, and its solutions. We then define the light-front form of the GPD for the Schwinger model, and discuss its kinematical structure and analytical properties.  In section~\ref{SEC4}, the boosted form of the qGPD 
is discussed for arbitrary velocities. We show that in the luminal case, the GPD is fully recovered at the non-perturbative level. In section~\ref{SEC5}, we present the discretized form of the Schwinger model Hamiltonian and boost operator. Using these results, in section~\ref{SEC6}, we briefly discuss the tensor network method used to simulate QED2 after the Jordan-Wigner transformation. We then present numerical results for qPDFs and qGPD for both a heavy and light mass system. We conclude in section~\ref{SEC7}. In the Appendices, we briefly recall the Poincare algebra used in the main text, and summarize the spin mapping of  QED2 on a 1D spatial lattice with open boundary conditions. In addition, we briefly comment on the structure of quantum algorithms to simulate qGPDs.

\section{The massive Schwinger model}
\label{SEC2} 

The massive Schwinger model, QED2 with fermions, is described by~\cite{Schwinger:1962tp,Coleman:1976uz}
\bea
\label{A1}
S=\int d^2x\,\bigg(\frac 14F^2_{\mu\nu} +\frac{\theta \tilde F}{2\pi}+\overline \psi (i\slashed{D}-m)\psi\bigg)
\eea
with the covariant derivative $\slashed{D}=\slashed{\partial}+ig\slashed{A}$. The bare fermion mass is $m$, and the coupling $g$ has mass dimension. At strong coupling  with $m/g\lesssim 0.3$, the spectrum is  composed of light mesons and baryons, with a C-even vacuum independent of $\theta$ 
\bea
m_\eta^2=m_S^2+m_\pi^2=\frac {g^2}\pi-\frac{m\langle \overline{\psi}\psi\rangle_0}{f^2},
\eea 
with $f=1/\sqrt{4\pi}$, the $\eta$ meson decay constant~\cite{Grieninger:2023ufa}. The vacuum chiral condensate is finite in the chiral limit with
$\langle\overline\psi \psi\rangle_0=-\frac{e^{\gamma_E}}{2\pi}m_S$
~\cite{Sachs:1991en,Steele:1994gf}, 
where $\gamma_E=0.577$ is the Euler constant. Hence
\bea
\label{WC1}
\frac{m_\eta}{m_S}=
\bigg(1+2e^{\gamma_E}\frac{m}{m_S}\bigg)^{\frac 12}\approx 
1+e^{\gamma_E}\frac m{m_S} \,.
\eea
 At weak coupling with $m/g\gtrsim 0.3$, the spectrum is composed of heavy mesons, with doubly degenerate C-even and C-odd vacuua at $\theta=\pi$. 
The $\eta$-mass is expected to asymptote
$m_\eta\rightarrow 2m$. Note that at weak coupling (\ref{WC1}) we have $e^{\gamma_E}\approx 1.78$ which is close to 2,
hence allowing a smooth and continuous transition in the eta mass from weak to strong coupling.

In time-like axial gauge $A^0=0$, the Hamiltonian reads
\bea
\label{H1}
\mathbb H=\int dx \bigg(\frac 12 
E^2+\frac{E\theta}\pi +
\overline\psi (i\gamma^x D^x +m)\psi\bigg)
\eea
with the electric field $E=-\partial^0A^x$. Here, $\theta/\pi$ plays the role of an external electric flux. The Gauss law constraint is fixed by the charged fermions
\bea
\partial_xE=gj^0\,.
\eea
Our conventions for the gamma matrices are
\bea
\gamma^0=
\begin{pmatrix}
1& 0\\
0&-1
\end{pmatrix}\qquad
\gamma^x=
\begin{pmatrix}
0& 1\\
-1&0
\end{pmatrix}.
\eea
The lattice version of the Hamiltonian in Eq.~(\ref{H1})  follows from the original staggered fermion formulation in~\cite{Banks:1975gq}. Its mapping to spin degrees of freedom will be given in section~\ref{SEC5} below.

\section{GPDs}
\label{SEC3}
The light-front Hamiltonian of QED$_2$ is very similar to that of QCD$_2$, apart from the $U(1)$ anomaly. In the massless case, QED$_2$ is exactly solvable and contains a single massive bosonic excitation with mass $m_S$. In contrast, the massive theory is not exactly solvable and its spectrum consists of multi-boson states.

\subsection{Light-front wave functions}

In the 2-particle Fock approximation, massless QED$_2$ exhibits a ground state and a tower of spurious states that disappear in the continuum when higher
Fock states contributions are included~\cite{Bergknoff:1976xr}. By contrast, massive QED$_2$ in the 2-particle Fock space approximation exhibits a ground state and a tower of genuine multi-meson states. Numerical studies indicate that the ground state is dominated by the two-particle Fock component, whereas the first excited state already receives a substantial four-particle contribution (a meson–meson bound state) in the strong coupling limit~\cite{Mo:1992sv}.

With this in mind, the light-front wavefunctions $\phi_n(\zeta)$ in the 2-particle Fock-space approximation follow from a Bethe-Salpeter-type derivation. More specifically, the
light-front wavefunctions solve~\cite{Bergknoff:1976xr,Grieninger:2024cdl}
\begin{align}
\label{TH1}
M_n^2\phi_n(x)=&m_S^2\int_{0}^1\,dy\,\phi_n(y)+\frac {m^2}{x(1-x)}\phi_n(x)\nonumber\\
&-m_S^2\,{\rm PP}\int_{0}^1\,dy\,\frac{\phi_n(y)-\phi_n(x)}{(x-y)^2}\,,
\end{align}
with $xP$  the momentum fraction of the parton. Here, PP is short for the principal part. Eq. (\ref{TH1}) is the 't Hooft equation~\cite{tHooft:1974pnl}, modulo the U(1) anomaly contribution (first term on the right-hand side). 
The longitudinal kinetic contribution (second term on the right-hand side)  is singular at the endpoints $x=0,1$,  forcing the light-front wavefunction to vanish at the edges.
Note that the 2-particle Fock-contribution is the leading contribution in the limit of a large number of colors in QCD$_2$. 

In the massless limit with $m\rightarrow 0$, the 
spectrum is given by a single massive boson with 
squared mass $M_0^2\rightarrow m_S^2$. In the massive limit with $m\gg m_S$, the solution is peaked around $x=\frac 12$, with $M_n\approx 2m$. In general, Eq.~(\ref{TH1}) admits a tower of excited states for finite $m/m_S$ since the PP part in (\ref{TH1}) is strictly confining. 
The lowest state parton distribution function is  of the form
\bea
\label{QM0}
q_\eta(x )=|\phi_0 (x )|^2.
\eea
Since the $\eta$ meson is flavor neutral, the 
fermion and anti-fermion distribution functions are identical.

\begin{figure}
    \centering
    \includegraphics[width=0.99\linewidth]{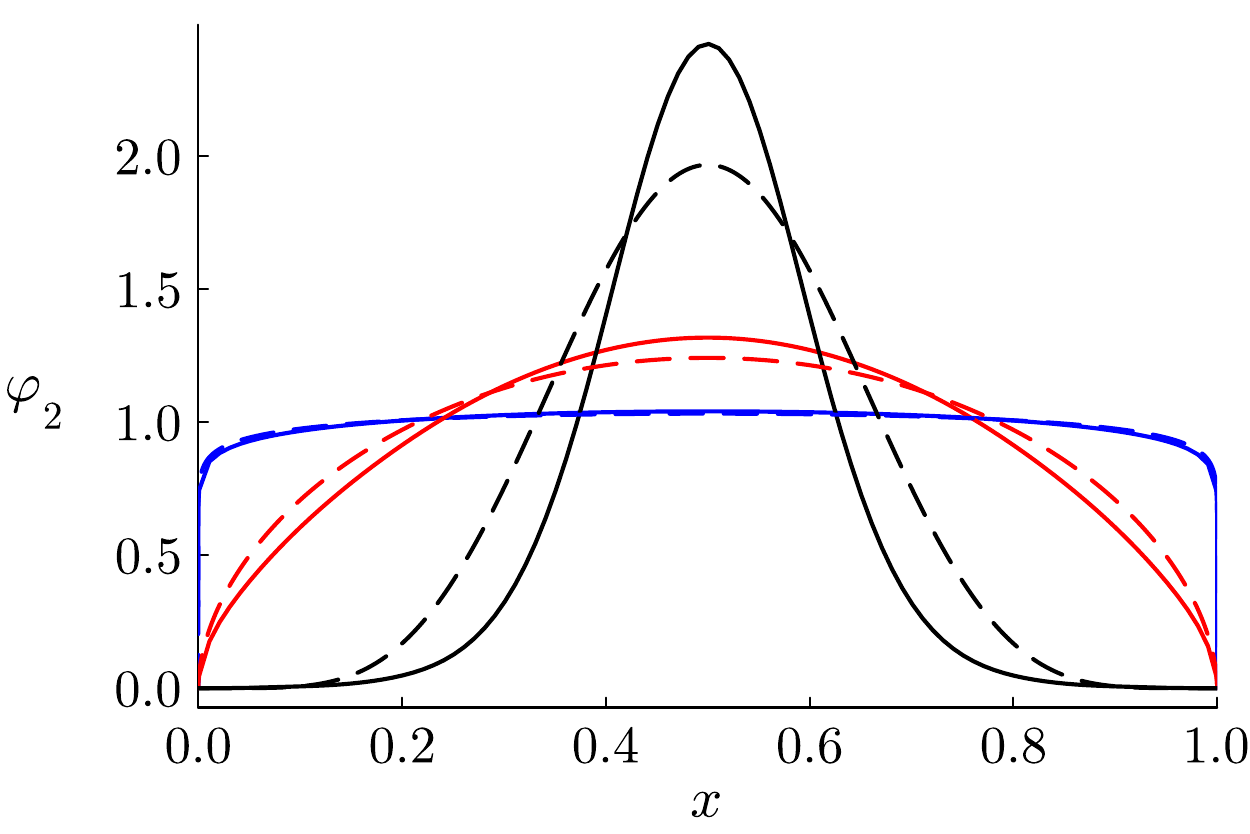}
    \caption{Solution to \eqref{TH1} for $\beta=0.1\sqrt{3}/\pi  $ (blue), $\beta=\sqrt{3}/\pi$ (red), $\beta=10\sqrt{3}/\pi  $ (black) using 13 Jacobi polynomials. See~\cite{Grieninger:2024cdl} for more details. The dashed lines are \eqref{ansatz} plotted for the same $\beta$ values.}
    \label{qfunct}
\end{figure}

The integral equation (\ref{TH1}) is a relative of the 't Hooft equation and can be solved numerically. For that, we use a direct discretization in terms of a matrix eigenvalue equation as for the 't Hooft equation~\cite{Zubov:2016bqs}. The solutions follow numerically by expanding in a complete basis. We solved the eigenvalue problem in~\cite{Grieninger:2024cdl}. See Appendix B in Ref.~\cite{Grieninger:2024cdl} and Ref~\cite{Mo:1992sv} for more details. In Fig.~\ref{qfunct} (solid curves), we show typical solutions in the strong and weak coupling regimes.
We will refer to these solutions as ``exact solution'' even though the solutions are approximated, in the sense that we cut the expansion in the Jacobi basis at a finite number of Jacobi polynomials. The coefficients of higher Jacobi polynomials are rapidly suppressed. 

For the ground state in the strong coupling regime and for comparison, we can use a normalized polynomial ansatz with enforced endpoint boundary conditions. We have
\bea
\phi_0(x)=\frac{x^{\beta}(1-x)^{\beta}}
{[\mathrm{B}(2\beta+1,\,2\beta+1)]^{\frac 12}}\,,
\label{ansatz}
\eea
where $B(a,b)$ is  Euler beta function. The end-point exponent $\beta$ is fixed by the small-$x$ analysis of (\ref{TH1})   $m^2_S -m^2= \pi\,\beta\,\cot(\pi\beta)\,$. In fig.~\ref{qfunct} (dashed curves), we compare the variational ansatz (\ref{ansatz}) to the numerically exact results. As evident from the figure, the variational ansatz works well in the strong coupling limit, i.e. at small $\beta$. The explicit form of the ansatz will prove useful for some of the analytical results to be developed below.

\subsection{Light-front GPDs}

In two dimensions, the light-front GPD for the $\eta$ meson in QED$_2$ is the off-diagonal matrix element of the analogue of the leading twist-2 quark operator in 4 dimensions,  given by
\begin{align}
\label{LFGPD}
&H_\Gamma(x,\xi,t)=
\int_{-\infty}^{+\infty} \frac{dz^-}{2\pi}\,
e^{ix P^+z^-} \\ \times &
\Big\langle P\!+\!\frac\Delta 2\Big|\overline\psi\Big(-\frac{z^-}{2}\Big)\,\Gamma\bigg[\!-\frac {z^-}2, \frac {z^-}2\bigg]_L\psi\Big(\frac{z^-}{2}\Big)\Big|P\!-\!\frac\Delta 2\Big\rangle\,,\nonumber
\end{align}
where $\big[x,y\big]_L = e^{-ig\int_x^y dz^- A^+(z)}$ is the Wilson line on the light front and $\Gamma$ is a Dirac operator.

\subsubsection{Kinematics}

Here, we denote the momenta of the initial and final $\eta$ meson as $p_1, p_2$ with mass $m_\eta=M_0$. We introduce the following notation
\bea
P^\mu=\frac{p_1^\mu+p_2^\mu}{2}\,,\qquad
\xi=\frac{p_1^+-p_2^+}{p_1^++p_2^+}\,,\nonumber\\
t=(p_2-p_1)^2=\Delta^2=-\frac{4\xi^2 M_0^2}{1-\xi^2}\,,
\label{eq:kinematics}
\eea
where $\xi$ is the skewness and $t$ the momentum transfer. In addition, we have $\Gamma=\gamma^+=\gamma^0+\gamma^1$, $\gamma^5=\gamma^0\gamma^1$,  with $\gamma^0=\sigma^3$ and $\gamma^1=i\sigma^2$. 
Both the in- and out-meson states are on-shell $(P\pm \Delta/2)^2=m_\eta^2$. 
Note that in 2-dimensions
$\overline\psi \gamma^+\psi=\overline\psi\gamma^+\gamma^5\psi$, as the
axial-vector is dual to the vector, and kinematically $P^0=P^+$. In addition, there is no transverse momentum in 2 dimensions. The skewness is tied to the momentum transfer through the mass-shell condition
\bea
\label{FIXED}
m_\eta^2=\frac t4\bigg(1-\frac 1{\xi^2}\bigg)\,.
\eea
As a result, we will drop $t$ in our notation from here on.

The full GPD splits into a sum of the DGLAP 
or $x>\xi$ (parton distribution regime) plus the ERBL or $x<\xi$
(form factor regime) contributions, which we write as
\bea
H_\Gamma(x,\xi,t)=H_>(x,\xi,t)+H_<(x,\xi,t)\,.
\label{FULLGPD}
\eea
While the tensor network analysis presented below, using the Kogut-Susskind formulation of QED$_2$, will be the main part of our numerical analysis, we now proceed in this section by analyzing both regions numerically and analytically using a semi-classical ansatz.

\subsubsection{DGLAP region: $x>\xi$}

Here, we provide a Fock space derivation of the light-front GPD in the DGLAP region
corresponding to $x>\xi$. For that, we set $z^+=0$ and $\Gamma=\gamma^+$ in the bilocal matrix element in \eqref{LFGPD}. The good component is $\psi_+$ in 
\begin{equation}
J^+=\bar\psi\gamma^+\psi\rightarrow 2:\!\psi_+^\dagger\psi_+\!:\,.
\end{equation}
with the mode expansion
\begin{align}
&\psi_+\Big(\frac {x^-}2\Big)=\\
&\int_0^\infty\!\frac{dk^+}{\sqrt{2\pi}}\,
\frac{1}{\sqrt{2k^+}}\Big[
b(k^+)\,e^{-i\frac{k^+}{2}x^-}
+d^\dagger(k^+)\,e^{+i\frac{k^+}{2}x^-}\Big].\nonumber
\end{align}
For the in/out states, we use the 2-Fock approximation
\bea
&&\hspace{-0.3cm}|p_1\rangle\!=\!\int_0^1\!\frac{dx_-\,\phi_0(x_-)}{\sqrt{x_-(1\!-\!x_-)}}\,
b^\dagger(x_-p_1^+)d^\dagger((1-x_-)p_1^+)|0\rangle,\nonumber\\
&&\hspace{-0.3cm}\langle p_2|\!=\!\langle 0|\!\int_0^1\!\frac{dx_+\,\phi_0^*(x_+)}{\sqrt{x_+(1-x_+)}}\,
d((1-x_+)p_2^+)b(x_+p_2^+)\,,\nonumber\\
\label{INOUT}
\eea
with $b^\dagger d^\dagger$ creating a quark-anti-quark pair in a boosted $\eta$ meson. Inserting the bilocal operator from \eqref{LFGPD} leads to the diagonal contraction $b^\dagger b$ (number-preserving) as the only contribution. The $z^-$ integral gives the phase delta, enforcing 
\begin{equation}
x P^+=\frac{k_1^+-k_2^+}{2}\quad\rightarrow\quad
x_\pm=\frac{x\pm \xi}{1\pm \xi}\,.
\end{equation}
Using the canonical anti-commutators and the spectator antiquark overlap, all plus-momentum factors cancel except for the Jacobian
\begin{equation}
\sqrt{\frac{1-\xi}{1+\xi}}=\sqrt{\frac{p_2^+}{p_1^+}}\,,
\end{equation}
leading to the following result for the GPD in the DGLAP regime
\bea
H_>(x,\xi,t)&=
\sqrt{\frac{1-\xi}{1+\xi}}\;
\phi_0(x_-)\,\phi_0(x_+)\,.
\label{eq:GPD}
\eea
The prefactor $\sqrt{\frac{1-\xi}{1+\xi}}$ originates from converting the longitudinal fraction $x$ defined with respect to $P^+$ in \eqref{LFGPD} to the fractions $x_-$ and $x_+$ defined with respect to $p_1^+$ (in state) and $p_2^+$  (out state) as in (\ref{INOUT}). 

\subsubsection{ERBL region: $x<\xi$}

In light-front time-ordered perturbation theory at $x^+=0$, pair creation by the \emph{good} current $J^+$ is absent. However, 
in the ERBL regime $0<x<\xi$, the contribution does not arise from instantaneous pair creation, but rather from the crossed-channel (lowest $t$-channel) meson pole. In two dimensions, bosonization implies that the gauge-invariant quark bilinear couples to the physical massive boson, generating a pole at $t=M_0^2\sim  m_\eta^2$ with a residue proportional to the distribution amplitude  
\be
H_{<}(x,\xi,t)\;\propto\;-\frac{\langle 0|\bar\psi\Gamma\psi|\eta\rangle\ \langle \eta|\bar\psi\Gamma\psi|0\rangle}{t-M_0^2+i0}\,
\varphi_0\!\Big(\frac{x}{\xi}\Big)\;. \hspace{10pt}
\ee
More specifically, within the lowest pole approximation, we find
\begin{align}
H_{<}(x,\xi,t)&=-\frac{\mathcal R_0(\xi)\;\varphi_0\!\left(\tfrac{x}{\xi}\right)}{t-M_0^2+i0}\,.
\label{eq:erbl1}
\end{align}
The (normalized to 1) distribution amplitude $\varphi_0$, which is distinct from $\phi_0$ by the normalization, readily follows
from the ansatz \eqref{ansatz}
\begin{equation}
\varphi_0(z)=\frac{z^\beta(1-z)^\beta}{\mathrm{B}(\beta+1,\beta+1)}\,,
\end{equation}
with the lowest pole residue $\mathcal R_0$ 
\begin{equation}
\mathcal R_0(\xi)=M_0^2\sqrt{1-\xi^2}\,.
\end{equation}
Since
\begin{equation}
t-M_0^2 = -M_0^2\,\frac{1+3\xi^2}{1-\xi^2}\,,
\end{equation}
we have
\begin{align}
H_<(x,\xi,t)=
\frac{\sqrt{1-\xi^2}}{B(\beta+1,\beta+1)}\,
\frac{1-\xi^2}{1+3\xi^2}\;
\Big(\tfrac{x}{\xi}\Big)^\beta\!\Big(1-\tfrac{x}{\xi}\Big)^\beta\,.
\label{eq:erbl2}\nonumber \\
\end{align}
Note that the more standard normalization of the distribution amplitude to $f_0=1/\sqrt{4\pi}$ (the 2D decay constant), should be followed by a rescaling of the residue $\mathcal R_0\rightarrow \mathcal R_0/f_0$, and would cancel out in the lowest pole contribution. This ensures the form factor sum rule of the GPD, as discussed below.

\begin{figure*}[t]
  \centering
  \includegraphics[width = 1.0\linewidth]{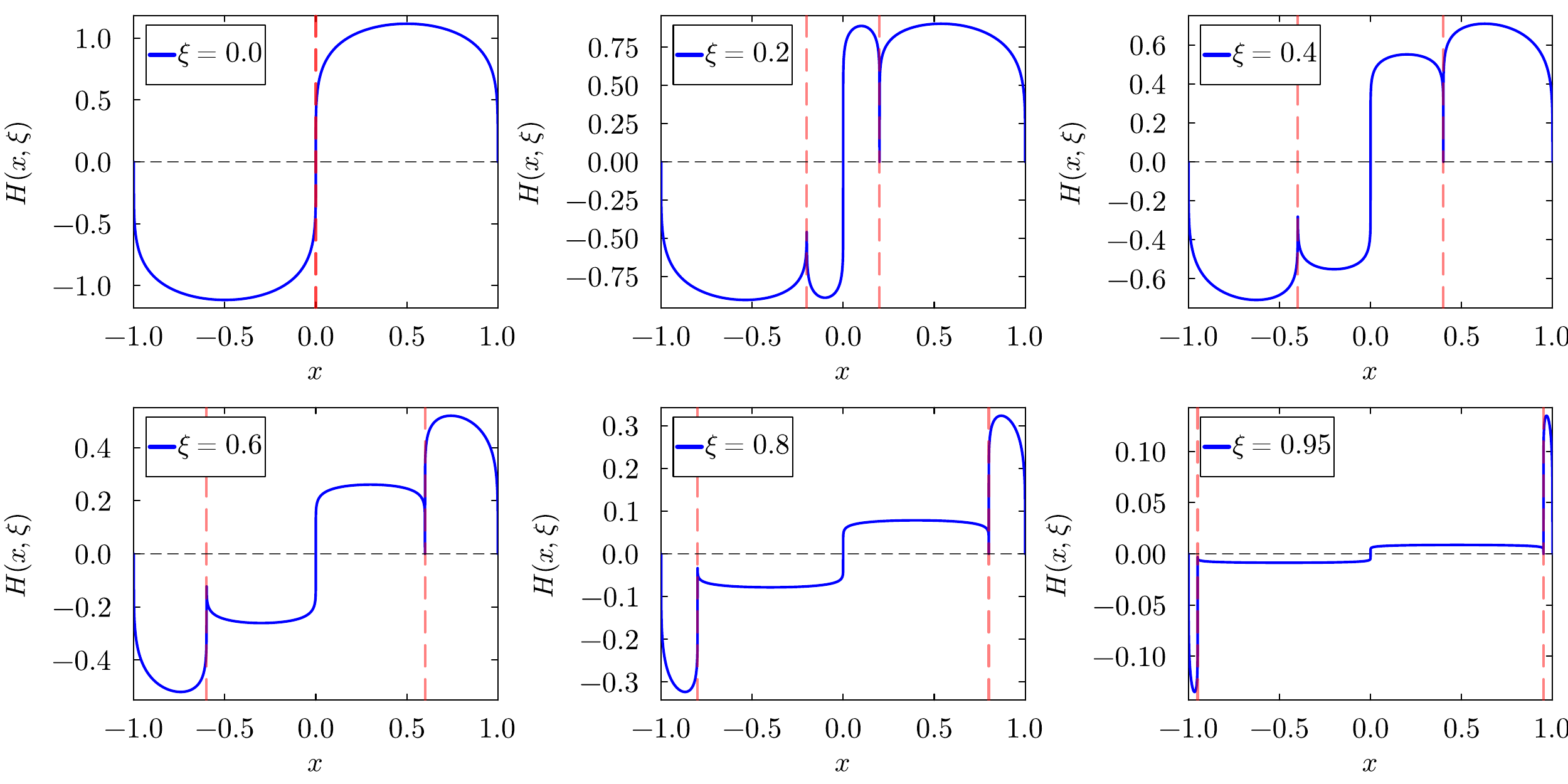}
  \captionsetup{justification=justified}
  \caption{Single pole GPD~\eqref{eq:fullgpd} for different values of the skewness $\xi$, as indicated in each panel, with $\beta=0.4$ and $M_0=1$. The red vertical lines delineating the DGLAP and ERBL regions are located at $x = \pm\xi$.}
    \label{fig:2part}
\end{figure*}

The final form of the lowest state and unpolarized $\eta$ meson GPD is
\begin{equation}
H(x,\xi,t)=
\begin{cases}
\text{Eq.\,\eqref{eq:GPD}}, & x>\xi,\\[0.2em]
\text{Eq.\,\eqref{eq:erbl2}}, & 0<x<\xi,\\[0.2em]
-\,H(-x,\xi,t), & x<0\,.
\end{cases}
\label{eq:fullgpd}
\end{equation}
for the quark and anti-quark. In Fig.~\ref{fig:2part} we show the behavior of the GPD~\eqref{eq:fullgpd} in the single pole approximation, for increasing values of the skewness  $\xi=0.0, 0.2, 0.4, 0.6, 0.8, 0.95$, with the index  $\beta=0.4$ (weak coupling) and a meson mass $M_0=1$ in units of $m_S$. In the 2-Fock approximation, the cusp at the null value of the GPD is clearly visible at $x=\pm \xi$, with a non-vanishing ERBL contribution from the single pole exchange in the $t$-channel.

\subsubsection{Properties}

In general, unpolarized GPDs in QCD obey generic relations, which can be extended to the 2D case. Time-reversal and hermiticity imply
\bea
H(x,\xi, t)=H^*(x, \xi, t)=H(x, -\xi, t)\,,
\eea
which make the spin-0 GPD real and  even in  skewness $\xi$. The GDP satisfies polynomiality, which translates to the fact that its $n$th Mellin moment is a polynomial of degree $\xi^n$, which is a consequence of Lorentz symmetry~\cite{Belitsky:2005qn}. 

In the DGLAP or PDF regime with $\xi<x<1$, the GDP  satisfies the positivity constraint (Cauchy-Schwarz inequality)~\cite{Pire:1998nw}
\bea
\label{BOUND}
|H_>(x,\xi)|\leq \bigg(\frac {1-\xi}{1+\xi}\bigg)^{\frac 12}\,(q(x_-)q(x_+))^{\frac 12}\,,
\eea
with the $\eta$ PDF in (\ref{QM0}). In the 2-Fock space approximation
and real light-front wave functions, the inequality saturates for
\eqref{eq:GPD}. 
The inclusion of higher Fock components weakens saturation.

\subsubsection{Continuity and cusp at $x=\xi$}

Near the matching point $x=\xi$, the two analytic forms join continuously but with discontinuous slopes.
From \eqref{eq:GPD} and \eqref{eq:erbl2}, we find the near-boundary expansions
\begin{align}
H_{<}(x,\xi,t)&\approx 
C_L\,\Big(1-\frac{x}{\xi}\Big)^{\beta}\,,\\
C_L(\xi,\beta)&=
\frac{\sqrt{1-\xi^2}}{\mathrm{B}(\beta+1,\beta+1)}\,
\frac{1-\xi^2}{1+3\xi^2}\,,
\end{align}
and
\begin{align}
&H_{>}(x,\xi,t)\approx 
C_R\,(x-\xi)^{\beta}\,,\\
&C_R(\xi,\beta)=
\frac{(2\xi)^{\beta}}{(1+\xi)^{2\beta}}\,
\sqrt{\frac{1-\xi}{1+\xi}}\;
[\mathrm{B}(2\beta+1,2\beta+1)]^{-1}.\nonumber\\
\end{align}
So $H$ is continuous and vanishes at $x=\xi$, but with discontinuous
derivatives (cusp) for $\beta<1$, which are given by
\begin{align}
&\partial_x H_{<}\sim -\frac{\beta\,C_L}{\xi}\Big(1-\frac{x}{\xi}\Big)^{\beta-1},\\
&\partial_x H_{>}\sim \beta\,C_R\,(x-\xi)^{\beta-1}\,.
\end{align}

\subsubsection{Reggeized ERBL regime}

In the ERBL regime, beyond the lowest pole, the crossed channel in 2D is a tower of meson states $\{\eta_n\}$ with
\begin{equation}
H_{<}(x,\xi,t)
=\sum_{n=0}^{\infty}\frac{\mathcal R_n(\xi)\,\varphi_n\!\big(\tfrac{x}{\xi}\big)}{M_n^2-t}\,,
\qquad z\equiv \frac{x}{\xi}\in(0,1).
\label{eq:ERBL-spectral}
\end{equation}
For large $n$, the light-front bound-state equation~\eqref{TH1} is semiclassical. A  WKB quantization of the longitudinal motion
produces an about linear squared-mass spectrum
\begin{equation}
M_n^2 \simeq M_*^2 + \lambda\,(n+\delta)\,,
\qquad n\gg 1,
\label{eq:WKB-linear}
\end{equation}
with slope $\lambda>0$ and offset $\delta$ determined by the short-distance
end-point index $\beta$ of the light-front wave function (and the anomaly term). Both $\lambda, M_*$ will be fixed below. 

The ERBL residues vary slowly with $n$.  
Their analytic continuation in the complex $j$-plane is
\begin{equation}
\mathcal R_n(\xi)\ \longrightarrow\ \beta(j,\xi),\qquad
\varphi_n(z)\ \longrightarrow\ \varphi(z;j)\,,
\end{equation}
with $\beta(j,\xi)$ and $\varphi(z;j)$ regular in a half-plane containing the positive real axis, and matching $\beta(n,\xi)=\mathcal R_n(\xi)$, $\varphi(z;n)=\varphi_n(z)$ at integer $j=n$.
With this in mind and using \eqref{eq:WKB-linear}, we can reorganize \eqref{eq:ERBL-spectral} as
\begin{align}
H_{<}(x,\xi,t)
&=\sum_{n=0}^\infty \frac{\beta(n,\xi)\,\varphi(z;n)}{M_*^2+\lambda(n+\delta)-t}\nonumber\\
&=\frac{1}{\lambda}\sum_{n=0}^\infty \frac{\beta(n,\xi)\,\varphi(z;n)}{n-\alpha(t)}\,,
\end{align}
where we introduced the Reggeized trajectory 
\begin{equation}
\alpha(t)\equiv \frac{t-M_*^2}{\lambda}-\delta.
\label{eq:alpha-def}
\end{equation}
The discrete and asymptotic sum can be unwound using the Sommerfeld--Watson transform with neutral signature
\begin{eqnarray}
&&\sum_{n=0}^\infty \frac{\beta(n,\xi)\,\varphi(z;n)}{n-\alpha}
=\nonumber\\
&&-\int_C \frac{dj}{2i\pi}\,\pi\cot(\pi j)\,
\frac{\beta(j,\xi)\,\varphi(z;j)}{j-\alpha}\,,
\end{eqnarray}
where $C$ is the Bromwich contour.

For $0<\xi<1$, the analytic factor $\beta(j,\xi)\!\propto\!(1-\xi^2)^{(2-j)/2}$ damps the integrand for ${\rm Re}\,j<0$, so that the contour $C$ is closed to the {\it left}. The deformation crosses the Regge pole at $j=\alpha(t)$, where the residue gives the leading contribution. All other singularities lie farther left in the $j$-plane and are thus subleading at large energy or small~$\xi$. Hence, the leading Regge term reads
\begin{eqnarray}
H_{<}^{\text{Regge}}(x,\xi,t)
=\frac{\pi}{\lambda}\,
\beta\big(\alpha(t),\xi\big)\,\varphi\big(z;\alpha(t)\big)\,\cot\!\big(\pi \alpha(t)\big)\nonumber\\
\label{eq:H-Regge-master}
\end{eqnarray}
modulo subleading cuts/poles.

In 2D, the ERBL $z$-shape remains controlled by the end-point index $\beta$ of the valence light-front wave function. We take
\begin{equation}
\varphi\big(z;j\big)=\frac{z^\beta(1-z)^\beta}{\mathrm B(\beta+1,\beta+1)}
\label{eq:DA-independent}
\end{equation}
independently of $j$ at leading WKB order, and parametrize the analytic residue as
\begin{equation}
\beta(j,\xi)=\beta_0\,(1-\xi^2)^{\frac{2-j}{2}}\,,
\label{eq:beta-analytic}
\end{equation}
which implements a positive signature (no alternating residues) and ensures damping for ${\rm Re}\,j<0$. The analytic continuation reduces to the first physical pole at $j=1$ (since $\alpha(M_0^2)=1$), reproducing the lowest-state limit and controlling the leading Regge behavior through the moving trajectory $\alpha(t)$. 

\begin{figure*}
    \centering
\includegraphics[width = 1.0\textwidth]{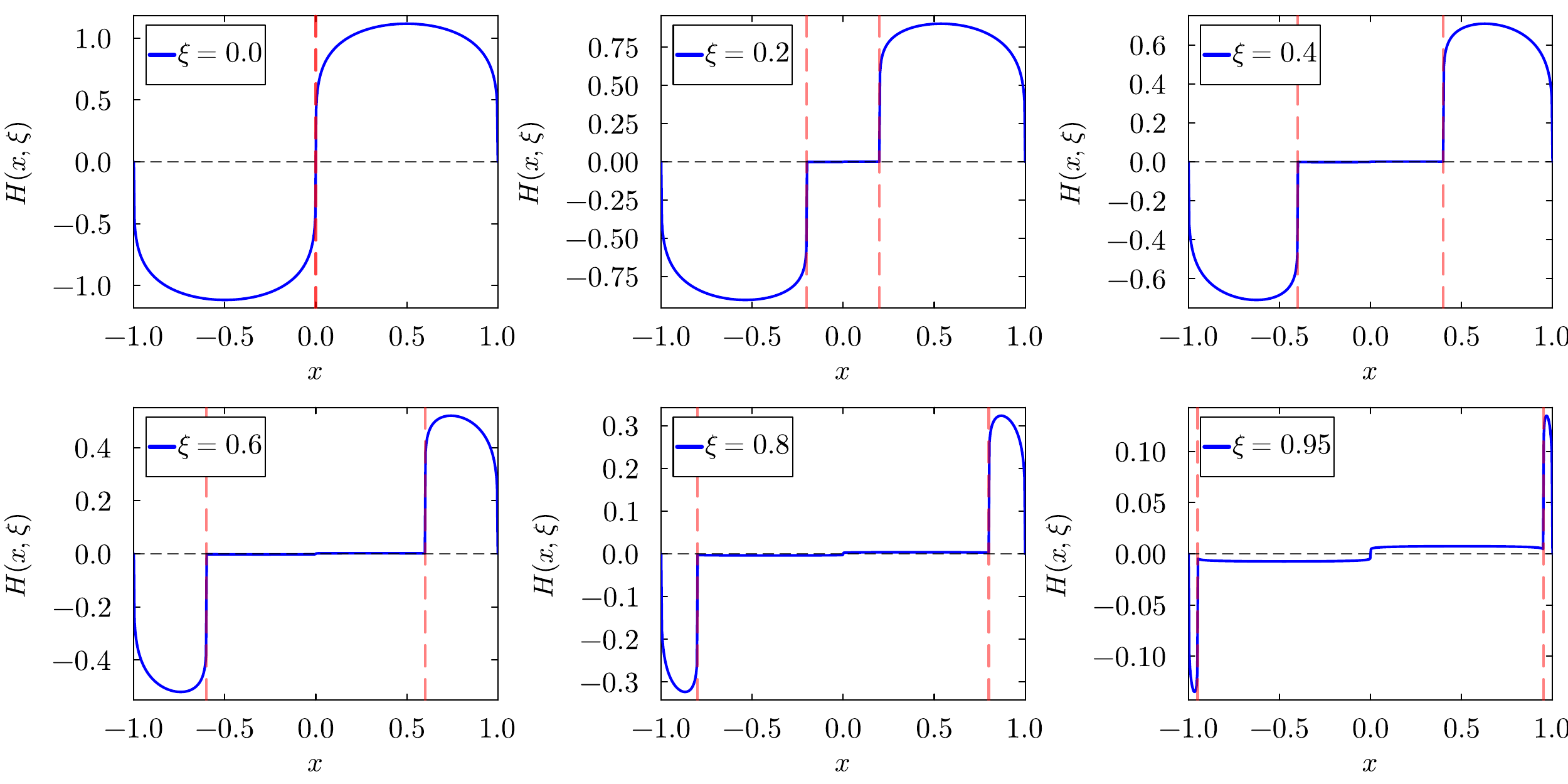}    
    \caption{Reggeized WKB GPD  for the same parameter choices as in Fig.~\ref{fig:2part}.}
   \label{fig:wkb}
\end{figure*}

Inserting \eqref{eq:DA-independent} and \eqref{eq:beta-analytic} into \eqref{eq:H-Regge-master}, yields the Reggeized ERBL result
\begin{align}
H_{<}^{\text{Regge}}(x,\xi,t)
=&\frac{\pi}{\lambda}\,
\beta_0\,\big(1-\xi^2\big)^{\frac{2-\alpha(t)}{2}}\nonumber \\
&\times\cot\!\big(\pi \alpha(t)\big)\;
\frac{\big(\tfrac{x}{\xi}\big)^\beta\big(1-\tfrac{x}{\xi}\big)^\beta}{\mathrm B(\beta+1,\beta+1)}\,,
\label{eq:ERBL-Regge-final}
\end{align}
for $0<x<\xi$. The choice $\mathcal{S}(\alpha)=\cot(\pi\alpha)$ is a signature-blind
real combination.  It has simple poles at all  integer $\alpha=J$
(rather than selecting only even or odd $J$).  
If, instead, one wishes to enforce explicit signature projection, we can use the replacements
\begin{eqnarray}
\mathcal{S}_{+}(\alpha)&=&+\tfrac12\cot\frac{\pi\alpha}{2}
\quad (\text{even }J)\,,\nonumber\\
\mathcal{S}_{-}(\alpha)&=&-\tfrac12\tan\frac{\pi\alpha}{2}\quad
(\text{odd }J)\,.
\end{eqnarray}
To fix  the overall normalization $\beta_0$ and the WKB slope $\lambda$, we match \eqref{eq:ERBL-Regge-final} to the lowest-pole result~\eqref{eq:erbl2} at $t\to M_0^2$. Near the ground-state pole, we have
\begin{eqnarray}
\alpha(t)&&=1+\alpha'(M_0^2)\,(t-M_0^2)+\cdots,\nonumber\\
\pi\cot\!\big(\pi\alpha(t)\big)&&=\frac{1}{\alpha'(M_0^2)}\;\frac{1}{\,t-M_0^2\,}+\,\cdots,
\end{eqnarray}
with $\alpha'(M_0^2)=1/\lambda$ for the linear trajectory \eqref{eq:WKB-linear}.
Thus \eqref{eq:ERBL-Regge-final} reduces to
\[
H_{<}^{\text{Regge}}(x,\xi,t)\ \xrightarrow[t\to M_0^2]{}\ 
\frac{\beta_0\sqrt{1-\xi^2}}{t-M_0^2}\;
\frac{\big(\tfrac{x}{\xi}\big)^\beta\big(1-\tfrac{x}{\xi}\big)^\beta}{\mathrm B(\beta+1,\beta+1)}\,,
\]
which must match the single pole form with residue $\mathcal R_0(\xi)=M_0^2\sqrt{1-\xi^2}$, hence $\beta_0=M_0^2$ and 
$\lambda={1}/{\alpha'(M_0^2)}$.
With this matching, the full Reggeized ERBL contribution is completely specified by the trajectory $\alpha(t)$ (slope $\alpha'$) and the end-point index $\beta$.

To fix the Regge parameters, we note that the confining integral in \eqref{TH1} acts as a nonlocal kernel. While this PP-kernel originates from the confining Coulomb law in 2D, we will use the harmonic approximation detailed in Appendix~\ref{app:harmonic} for a WKB estimate. Indeed, 
for slowly-varying $\phi(\zeta)$, the harmonic approximation gives\footnote{``Harmonic” here refers to the quadratic truncation of the nonlocal kernel around $\zeta=0$. 
Higher derivatives of $\phi$ and boundary effects generate subleading corrections and do not define a full local Hamiltonian.}
\begin{align}
&m_S^2\,{\rm PP}\!\int_{-1}^{1}\!d\zeta'\,
\frac{\phi(\zeta')-\phi(\zeta)}{(\zeta'-\zeta)^2} \nonumber\\
\;\approx\; &
\,m_S^2\,\partial_\zeta^2\phi(\zeta)
+\tfrac{\pi^2}{6}\,m_S^2\,\zeta^2\,\phi(\zeta)+\ldots
\end{align}
We use this quadratic approximation only to estimate the semiclassical level spacing. 
Treating the quadratic piece as the local potential in a Schr\"odinger-like WKB estimate with frequency 
$\omega\simeq \pi m_S/\sqrt{3}$, yields an approximately linear spectrum 
$M_n^2\simeq M_0^2+2\pi m_S^2 n$ up to corrections from higher gradients and the anomaly term. 
Consequently, the Regge slope in \eqref{eq:WKB-linear} is 
$\lambda\simeq 2\pi m_S^2$ with  $\alpha'(t)=1/\lambda\simeq 1/(2\pi m_S^2)$, and $M_*=M_0$.

In Fig.~\ref{fig:wkb}, we show the behavior of the GPD~\eqref{eq:fullgpd} 
following from the Reggeized WKB approximation~\eqref{eq:ERBL-Regge-final}, for increasing skewness  $\xi=0.0, 0.2, 0.4, 0.6, 0.8, 0.95$, with the index  $\beta=0.4$ (weak coupling) and a meson mass $M_0=1$ in units of $m_S$. In the WKB  approximation, the ERBL region is depleted after Reggeization, with
a cusp at $x=\pm \xi$, which is in contrast to the single pole approximation discussed earlier. The Reggeized results in Fig.~\ref{fig:wkb} for the anomalous $\eta$ in QED$_2$ at strong coupling are very similar to the light pion results in QCD$_2$~\cite{Burkardt:2000uu}, building on the results developed earlier in~\cite{Einhorn:1976uz}. This shows that the longitudinal features of QED$_2$ and QCD$_2$ are similar across $x=\xi$, with the duality between form factors and parton densities upheld. Recall that the results in QCD$_2$ are based on the exact and leading large $N_c$ contributions, while those developed here in QED$_2$ rely on the 2-Fock space approximation (DGLAP region) and the Reggeized limit (ERBL region),  with the lowest state fixed by the U(1) anomaly.

\subsubsection{Form factors}

The transition matrix elements of the local current $J^+$ evaluated between meson states gives the form factors
\bea
\langle m,P'|\,J^+(0)\,|n,P\rangle&=&(P'^++P^+)\,F_{mn}(t)\nonumber\\
&=&2P^+\,F_{mn}(t).
\label{eq:ffJ+}
\eea
Since in 2D, we have $P^+=P^0$, kinematically, the form factor in (\ref{eq:ffJ+}) holds for $J^0$ as well by covariance. In the 2-body Fock space approximation, we have
\bea
F_{mn}(t)
=\sqrt{\frac{1-\xi}{1+\xi}}\int_{\xi}^{1}dx\;\phi_m(x_-)\,\phi_n(x_+).
\eea
Since $J^+$ is the good current, there is no pair-creation contribution at $x^+=0$. All the dynamics are in the light-front wave functions $\phi_n$, which include anomaly effects via \eqref{TH1}. The diagonal or elastic form factor $F(t)=F_{00}(t)$ for the lowest state, obeys the sum rule 
\begin{equation}
\int_{-1}^1 dx\,H(x,\xi,t)=F(t)
\end{equation}
where both the quark and antiquark GPDs are added. Using the GPD results for the
DGLAP region, we have
\begin{align}
F_{q,>}(t)
=&\,\sqrt{\frac{1-\xi}{1+\xi}}\;\mathcal N^2\,(1-\xi^2)^{-2\beta}
\nonumber\\
&\,\times\int_{\xi}^{1}\!dx\;\big[(x^2-\xi^2)(1-x)^2\big]^{\beta}\nonumber\\
=&\,\sqrt{\frac{1-\xi}{1+\xi}}\;
\mathcal N^2\,
(1-\xi)^{1+3\beta}\,(1+\xi)^{-2\beta}\,
(2\xi)^{\beta}\nonumber\\&\,\times 
\mathrm{B}(\beta\!+\!1,\,2\beta\!+\!1)\;
\nonumber\\&\,\times
{}_2F_{1}\!\left(\!-\,\beta,\ \beta\!+\!1;\ 3\beta\!+\!2;\frac{\xi-1}{2\xi}\right),
\label{eq:FqD}
\end{align}
with normalization $1/\mathcal N^{2}=\mathrm{B}(2\beta+1,\,2\beta+1)$. The hypergeometric function in \eqref{eq:FqD} is obtained after integration over the DGLAP kinematics. For that, it is useful to redefine the integration variable
$x=\xi+(1-\xi)u$ with $u\in[0,1]$, so that $dx=(1-\xi)du$, $1-x=(1-\xi)(1-u)$, and $x^2-\xi^2=(1-\xi)u\,[2\xi+(1-\xi)u]$. The remaining integral is of Euler type
\bea
&&\int_0^1 du\,u^{a-1}(1-u)^{c-a-1}(1+ku)^{-b}\nonumber\\
&&=\mathrm{B}(a,c-a)\,{}_2F_1(a,b;c;-k),
\eea
with $a=\beta+1$, $c=3\beta+2$, $b=-\beta$, $k=(1-\xi)/(2\xi)$, giving \eqref{eq:FqD}. 

For the ERBL region, we have
\begin{eqnarray}
F_{q,<}(t)&&=\int_{0}^{\xi}\!dx\,H_{<}(x,\xi,t)\nonumber\\
&&=\frac{M_0^2\sqrt{1-\xi^2}}{M_0^2-t}\,
\int_{0}^{\xi}\!dx\;
\frac{(\frac{x}{\xi})^\beta(1-\frac{x}{\xi})^\beta}{\mathrm{B}(\beta+1,\beta+1)}\,.\nonumber\\
\end{eqnarray}
The last integral can be done using
\begin{equation}
  \int_{0}^{\xi}\!\!dx\;\frac{(\frac{x}{\xi})^\beta(1-\frac{x}{\xi})^\beta}{\mathrm{B}(\beta\!+\!1,\beta\!+\!1)}
=\xi\int_0^1\!\!dz\;\frac{z^\beta(1-z)^\beta}{\mathrm{B}(\beta\!+\!1,\beta\!+\!1)}=\xi, 
\end{equation}
which results in 
\begin{equation}
   F_{q,<}(t)=\frac{M_0^2}{M_0^2-t}\,\sqrt{1-\xi^2}\;\xi
=\frac{(1-\xi^2)^{3/2}}{1+3\xi^2}\;\xi \,.
\label{eq:FqE}
\end{equation}
The  net quark form factor is $F_q(t)=F_{q,>}(t)+F_{q,<}(t)$. For a neutral meson such as the $\eta$, one combines the quark and the antiquark charges as
\begin{align}
F(t)=&\,
e_q\!\int_{0}^1\!dx\,H(x,\xi,t) 
+e_{\bar q}\!\int_{-1}^0\!dx\,(-H(x,\xi,t)) \nonumber \\
=&\,e\!\int_{-1}^1\!dx\,H(x,\xi,t)=0
\,.
\end{align}
which vanishes, since the GPD is $x$-odd.  

For completeness, we note that the Reggeized form factor follows from integrating \eqref{eq:ERBL-Regge-final} over $x\in(0,\xi)$,
\begin{align}
&F_{q,<}^{\text{Regge}}(t)
=\int_{0}^{\xi}\!dx\,H_{<}^{\text{Regge}}(x,\xi,t)\nonumber \\
&=
\frac{\pi}{\lambda}\,\beta_0\,
\cot\!\big(\pi\alpha(t)\big)\;
\Big(\frac{4M_0^2}{\,4M_0^2-t\,}\Big)^{\!\frac{2-\alpha(t)}{2}}
\sqrt{\frac{t}{\,t-4M_0^2\,}}\,,
\label{eq:Fq-ERBL-Regge}
\end{align}
after using the 2D on-shell condition. 
As $t\to -\infty$ (fixed $\alpha(t)$), (\ref{eq:Fq-ERBL-Regge}) asymptotes to
\begin{equation}
F_{q,<}^{\rm Regge}(t)\ \sim\
\cot\!\big(\pi\alpha(t)\big)\;|t|^{-\frac{\alpha(t)-1}{2}},
\end{equation}
with manifest Regge behavior\footnote{The shift in the exponent arises because the ERBL form factor is a kinematically reduced matrix element, not a full invariant amplitude.}. Note that 
the  single pole limit follows from  $t\to M_0^2$ with $\sin\pi\alpha \sim \pi\alpha'(M_0^2)(t-M_0^2)$ and $\lambda=1/\alpha'$, 
\begin{equation}
\frac{M_0^2}{\,M_0^2-t\,}\,\sqrt{1-\xi^2}\;\xi\,.
\end{equation}
The electromagnetic form factor of a $C$-even meson vanishes, since the vector current is $C$-odd. Note that if instead the probe is a $C$-\emph{even} current such as the gravitational current or a flavor non-singlet vector current, which does not enforce $H(-x)=-H(x)$, then a \emph{neutral} target can have $F(0)=0$ but $F(t)\neq0$ for $t\neq0$. 

\section{Quasi-GPDs}
\label{SEC4}

To quantify the partonic content of the lowest meson $\eta$ in massive QED$_2$, we follow Ji’s quasi-distribution framework~\cite{Ji:2013dva} for boosted states. This formulation enables a controlled study of the continuum limit with tensor networks and allows us to investigate the convergence of quasi-distributions as the boost approaches the light cone.

We begin by introducing the rapidities of the meson states, which can be expressed in terms of the skewness. For a symmetric light-front momentum splitting between the incoming and outgoing states in the definition of the GPD, we have
\bea
p_{1,2}^\pm=P^\pm\mp \frac{\Delta^\pm}2=P^\pm(1\mp \xi)\,.
\eea
The rapidities of the in- and outgoing states follow as
\bea
\chi_{1,2}=\frac 12{\rm ln}\bigg(\frac{p_{1,2}^+}{p_{1,2}^-}\bigg)=
{\rm ln}\bigg(\frac{p_{1,2}^+}{m_\eta}\bigg)\,,
\eea
where we used the on-shell conditions $p_{1,2}^+p_{1,2}^-=m_\eta^2$. The rapidity difference between the in- and out states is
\bea
\Delta\chi={\rm ln}\bigg(\frac{1+\xi}{1-\xi}\bigg)\,.
\eea
For the equal time frame appropriate for the qGPD, the on-shell conditions are
\bea
\label{TKIN}
p^z_{1,2}&=&P(1\mp \xi)=\gamma(v_{1,2})\,m_\eta\,v_{1,2}\,,\nonumber\\
p^0_{1,2}&=&(p^{z2}_{1,2}+m_\eta^2)^{\frac 12}=\gamma(v_{1,2})\,m_\eta\,.
\eea
Here, $P$ corresponds to the luminal limit $v_{1,2}\rightarrow 1$.
The rapidities follow from (\ref{TKIN}) and are given by
\begin{align}
\chi_{1,2}\!=&\,\frac 12{\rm ln}\bigg(\frac{1+v_{1,2}}{1-v_{1,2}}\bigg)\nonumber\\
=&\,\frac 12{\rm ln}\bigg(
\frac{(m_\eta ^2+P^2(1\mp\xi)^2)^{\frac 12}+P(1\mp \xi)}
{(m_\eta ^2+P^2(1\mp\xi)^2)^{\frac 12}-P(1\mp \xi)}\bigg)\nonumber\\
=&\,\ln\!\left(\frac{P (1\mp \xi)+\sqrt{m_\eta^2+P^2(1\mp\xi)^2}}{m_\eta}\right)\,,
\end{align}
or equivalently
\begin{align} \label{eq:GPD_phases}
\chi_1=&\,{\rm sinh}^{-1}\bigg(\frac {P(1-\xi)}{m_\eta }\bigg)\,,\nonumber\\ 
\chi_2=&\,{\rm sinh}^{-1}\bigg(\frac {P(1+\xi)}{m_\eta }\bigg)\,.
\end{align}
In this case, the squared momentum transfer is related to the skewness through
\begin{align}
\label{BFIXED}
m_\eta=&\frac t2\!-\!(1-\xi^2)P^{2}
\!+\!\bigg(\!(m_\eta^2+(1+\xi^2)P^2)^2\!-\!4\xi^2P^{4}\!\bigg)^{\!\frac 12}
\end{align}
which reduces to (\ref{eq:kinematics}) in the large  $P$-limit. The  partonic qGPD  of a boosted $\eta$ meson is defined as
\begin{align}
\label{QGPD}
 H_\Gamma(x,\xi, v)= &\,
\int_{-\infty}^{+\infty} \frac{dz}{2\pi}\,
e^{- izx P} \nonumber\\
&\hspace*{-2.5cm}\,\times \langle \eta|\,e^{- i\chi_2 \mathbb K}\,\overline\psi(0,- z)[- z,+ z]_S\Gamma_v\psi(0,+ z)\,e^{i\chi_1 \mathbb K}\,|\eta\rangle \,.
\end{align}
Here, $\Gamma_v=\gamma^0$ (unpolarized), $\Gamma_v=\gamma^1$ (polarized),
$\Gamma_v={\bf 1}$ (scalar) and  $\Gamma_v=\gamma^5$ (axial).
The Wilson line  $[x,y]_S$ is taken along the spatial direction. For convenience, we define the non-local and bilinear qGPD operator 
\bea
{\cal O}_\Gamma(-z, z)=\overline\psi(0,- z)[- z,+ z]_S\Gamma_v\psi(0,+ z)\,.
\eea
In Appendix~\ref{app:discsym}, we discuss the symmetries of the matrix element. The GPD is obtained from (\ref{QGPD}) in the large- momentum limit. By analogy
with the 4D case, we can also define $\Gamma_v=\gamma^0+v\gamma^1$ with
\begin{equation}
    v={\rm tanh}\frac 12(\chi_1+\chi_2)\,.
\end{equation}

\section{The Schwinger model on a spatial lattice}
\label{SEC5}

We discretize the Schwinger model using a lattice of length $L=Na$, where $a$ is the lattice spacing and $N$ the number of staggered fermion sites. The fermion field $\varphi_n$ is placed on each site $n=0,..,N-1$. The sites are connected by gauge links $U_{n,n+1}=e^{-iagA_x(n)}$.

\subsection{Hamiltonian}

The staggered lattice Hamiltonian 
for $\theta=0$ is then given by
\begin{align}
\mathbb H=
&\frac{1}{2a}\sum_{n=0}^{N-2}
\bigg(
\varphi_{n+1}^\dagger
iU_{n+1,n}\varphi_n-
\varphi_n^\dagger
iU_{n,n+1}\varphi_{n+1}
\bigg)\nonumber\\
&+m\sum_{n=0}^{N-1}(-1)^n
\varphi_n^\dagger\varphi_n+\frac {a}2\sum_{n=0}^{N-2}E^2_n\,.
\end{align}
In the compact U(1) formulation, we can identify the gauge link with an angular variable $\theta_n=-agA^x_n$, and the electric field by its canonically conjugate orbital momentum, $E_n=gL_n$,
\bea
[\theta_m, L_m]=i\delta_{mn}.
\eea
This implies that the electric fluxes in an open chain are quantized, $L_n=0,\pm 1, ...$. 

In the Kogut-Susskind Hamiltonian formulation, the staggered fermions are mapped onto spatial lattice sites by assigning the upper component to even sites and the lower component to odd sites (or vice versa), 
\bea
\psi(0,z=na)\!=\!\frac 1{\sqrt a}\!
\begin{pmatrix}
   \psi_e(n) \\
   \psi_o(n)
\end{pmatrix}
\!=\!\frac 1{\sqrt a}\!
\begin{pmatrix}
   \varphi_{n: {\rm even}} \\
   \varphi_{n+1: {\rm odd}}
\end{pmatrix}\,,\;\;
\eea
with $0\leq n\leq N-1$. 
The staggered lattice Hamiltonian can be mapped on  a spin Hamiltonian using a Jordan-Wigner transformation~\cite{Jordan:1928wi,Kogut:1974ag,Susskind:1976jm,Banks:1975gq}
\bea\label{eq:JW}
\varphi_n&=&\frac 12\prod_{m<n} [iZ_m] (X_n-iY_n)\,,\nonumber\\
\varphi^\dagger_n&=&
\frac 12\prod_{m<n}[-i Z_m] (X_n+iY_n)\,.
\eea
This leads to the following result for the discretized Hamiltonian
\begin{align}
\label{eq:Ham}
\mathbb H=& \frac{1}{8a}\sum_{n=0}^{N-2}\Big((X_{n+1}+iY_{n+1}) e^{-i\theta_n}(X_{n}-iY_{n})+{\rm h.c.}\Big)\nonumber\\ 
&+\frac{m}{2}\sum_{n=0}^{N-1}(-1)^n Z_n
-\frac {Nm}2+\frac{a g^2}{2}\sum_{n=0}^{N-2}L^2_n\,,
\end{align}
where $L_n$ is the local electric flux field operator 
\begin{equation}
\label{GAUSS}
    L_n=L_0+\sum_{m=0}^n\frac{Z_m+(-1)^m}{2}\,.
\end{equation}
The constant $-Nm/2$ in Eq.~(\ref{eq:Ham}) normalizes the ground state energy to zero at strong coupling. The angular variable $\theta_n=-agA_n^x$ 
can be eliminated by the unitary redefinition of the spin matrices
\bea
\frac 12(X_n-iY_n)\rightarrow 
\bigg(\prod_{0<m<n}e^{-i\theta_m}\bigg)\,\frac 12(X_n-iY_n)\,,\;\;
\eea
which does not affect the commutation rules modulo the left endpoint. This freedom is traced back to the leftover residual and time-independent gauge transformation of the spatial vector field in the temporal (Weyl) gauge. The final form of the Hamiltonian is
\begin{align}
\label{eq:HamX}
\mathbb H=&\,\frac{a g^2}{2}\sum_{n=0}^{N-2}L^2_n
 +\frac{1}{4a}\sum_{n=0}^{N-1}\left(X_{n+1}X_n +Y_{n+1}Y_{n}\right)\nonumber \\
&+\frac{m}{2}\sum_{n=0}^{N-1}(-1)^n Z_n
-\frac {Nm}2 \;.
\end{align}

\subsection{Boost and momentum}

Next, we consider the momentum and boost operators, starting from the continuum stress–energy tensor $T^{\mu\nu}$. The Hamiltonian in Eq.~(\ref{H1}) is obtained from
\begin{equation}
    \mathbb H= \int dx^1 T^{00}(x^0, x^1)\,.
\end{equation}
Analogously, the boost operator is
\begin{equation}
    \mathbb K = \int dx^1 \left(x^0 T^{01} - x^1T^{00} \right)\,.
\end{equation}
On the $x^0=0$ surface, this reduces to
\begin{equation}
\mathbb K = - \int dx^1 x^1 T^{00}\,, 
\end{equation}
which is the form that we use in this work. After discretization, following the same steps used for the Hamiltonian, we obtain
\begin{align}
\mathbb{K}=&\,- a\sum_{n=0}^{N-1}(n-(N-1)/2) T^{00}_n;\nonumber\\
      T^{00}_n
      =&\,
      \frac12\left( \frac{1}{4a} (K_n+K_{n-1})\right)+\frac{m}{2}\left((-1)^n Z_n+1\right)\nonumber\\
      &\,+ \frac12\left(\frac{ag^2}{2}\left(L_n^2+L_{n-1}^2\right) \right),
\end{align}
Here, the kinetic terms are 
$K_n=X_{n+1}X_n+Y_{n+1}Y_n$ and, with open boundary conditions, $K_{-1}=K_{N-1}=L_{-1}=L_{N-1}=0$. The use of the lattice operator $T^{00}_n$, as opposed to the terms in the lattice Hamiltonian (\ref{eq:HamX}), ensures that the quasi-GPD maintains relevant symmetries, as discussed in Appendix \ref{app:discsym}. 

We now turn to the momentum operator $\mathbb{P}$. The total momentum involves a lattice translation by $2a$ between even-even or odd-odd sites~\cite{Janik:2025bbz}. The continuum expression is given by
\bea
\label{PX}
&&\mathbb P= \int dx^1 T^{01} = \int dx\, \psi^\dagger(0,x) \frac{D_x}i\psi (0,x)\,.\qquad
\eea
A lattice discretization analogous to the Hamiltonian and boost gives
\begin{align}
\label{Pn}
\mathbb{P}= &\,\sum_{n=0}^{N-2}T_n^{01}\\
T_n^{01} =&\,\frac{-i}{8a}\left((X_{n+2}+iY_{n+2})\,Z_{n+1}\, (X_{n}-iY_{n})-{\rm h.c.}\right)\nonumber\\
=&\,\frac 1{4a}\big(Y_{n+2}Z_{n+1}X_n-X_{n+2}Z_{n+1}Y_n\big).
\end{align}

\subsection{Quasi-GPDs}

Now we turn to the lattice discretized version of the quasi-GPD introduced above. It can be written in terms of the following discrete Fourier transform
\begin{equation}\label{eq:qGPD_staggered}
    H( x, \xi, \chi) = \frac{1}{\pi} \sum_{n=0,2,4,\dots}^{N-2} e^{i 2a (n-\frac{N}{2}+1) x P^1(\chi)} H( n, \xi, \chi)
\end{equation}
For the choice $\Gamma=\gamma^0$, and using staggered fermions, we obtain
\begin{equation}\label{eq:qGPD_staggeredII}
\begin{split}
    &H( x, \xi, \chi) =  \frac{1}{\pi}\sum_{n=0,2,4,\dots}^{N-2} e^{i 2a (n-\frac{N}{2}+1) x P^1(\chi)} \\
    &\times \bigg[ 
    \bra{\eta(\chi_2(\xi,\chi))} \sum_{i=0}^1 \varphi^\dagger_{N-n+i-1}\,\varphi_{n+i} \ket{\eta(\chi_1(\xi,\chi))} \\
    &\; -\;
    \bra{\Omega(\chi_2(\xi,\chi))} \sum_{i=0}^1 \varphi^\dagger_{N-n+i-1}\,\varphi_{n+i} \ket{\Omega(\chi_1(\xi,\chi))} 
    \bigg]\,,
\end{split}
\end{equation}
where the sum runs over the physical lattice sites. The momentum of the boosted $\eta$ meson is given by $P^1(\chi) = m_\eta \sinh(\chi)$, see Eq.~\eqref{eq:cont_disp}. Here, $\ket{\eta(\chi)}$ and $\ket{\Omega(\chi)}$ are the boosted $\eta$ state and ground state, respectively. The rapidities $\chi_1$ and $\chi_2$ are given in Eq.~(\ref{eq:GPD_phases}). 

The relevant operator insertions can be mapped to spin degrees of freedom using the Jordan-Wigner transformation in Eq.~(\ref{eq:JW}). For $n+i > N-n+i-1$,
\begin{equation}\label{eq:JW_op_insertion}
\begin{split}
    \varphi^\dagger_{N-n+i-1}\varphi_{n+i} = (-1)^{n-\frac{N}{2}-1}\sigma^+_{N-n+i-1}\\
    \times \left( \prod_{m = N-n+i-1  }^{n+i} Z_m\right)\sigma^-_{n +i} \;\;.
\end{split} 
\end{equation}
For the case where $n+i < N-n+i-1$, the bounds on the product of $Z$ are swapped. The quasi-PDF, denoted $q_\eta(x,\chi)$, can be obtained from this formula by taking $\xi = 0$, where~(\ref{eq:GPD_phases}) reduces to $\chi_{1/2}(\xi=0,\chi) = \chi$.

\section{Tensor network results}
\label{SEC6}

In this section, we present tensor network calculations for the state preparation, the relativistic energy-momentum dispersion relation, the qPDF, and the qGPD, each for two different masses:  $m/g =0.2$ and $m/g = 1$. We use a spatial lattice with $400$ fermion sites,  corresponding to $200$ physical sites, a lattice spacing of $a= 0.075$, and a coupling of $g= 0.5$. For the qPDF and qGPD, we take $\Gamma = \gamma^0$, and drop the subscript from here on. All reported tensor network calculations utilized software from the \texttt{ITensor} ecosystem \cite{itensor, ITensor-r0.3} in \texttt{Julia}~\cite{bezanson2014julia}. 

\subsection{State preparation}

\begin{table}[]
    \centering
    \begin{tabular}{c|c|c}
        $\delta E_\psi /m_\eta$ &   $m/g= 0.2$ & $m/g= 1.0$ \\ \hline 
        $\ket\Omega$& $2.531 \cdot 10^{-4}$ &  $2.198 \cdot 10^{-4}$   \\
        $\ket\eta$ & $3.749 \cdot 10^{-4}$ & $2.133 \cdot 10^{-4}$
    \end{tabular}
    \caption{The variances of the lattice Hamiltonian for the vacuum and $\eta$ states, for the considered mass parameters divided by the calculated mass gap of the $\eta$ state.}
    \label{tab:variances}
\end{table}

\begin{figure*}
    \includegraphics[width=.475\linewidth]{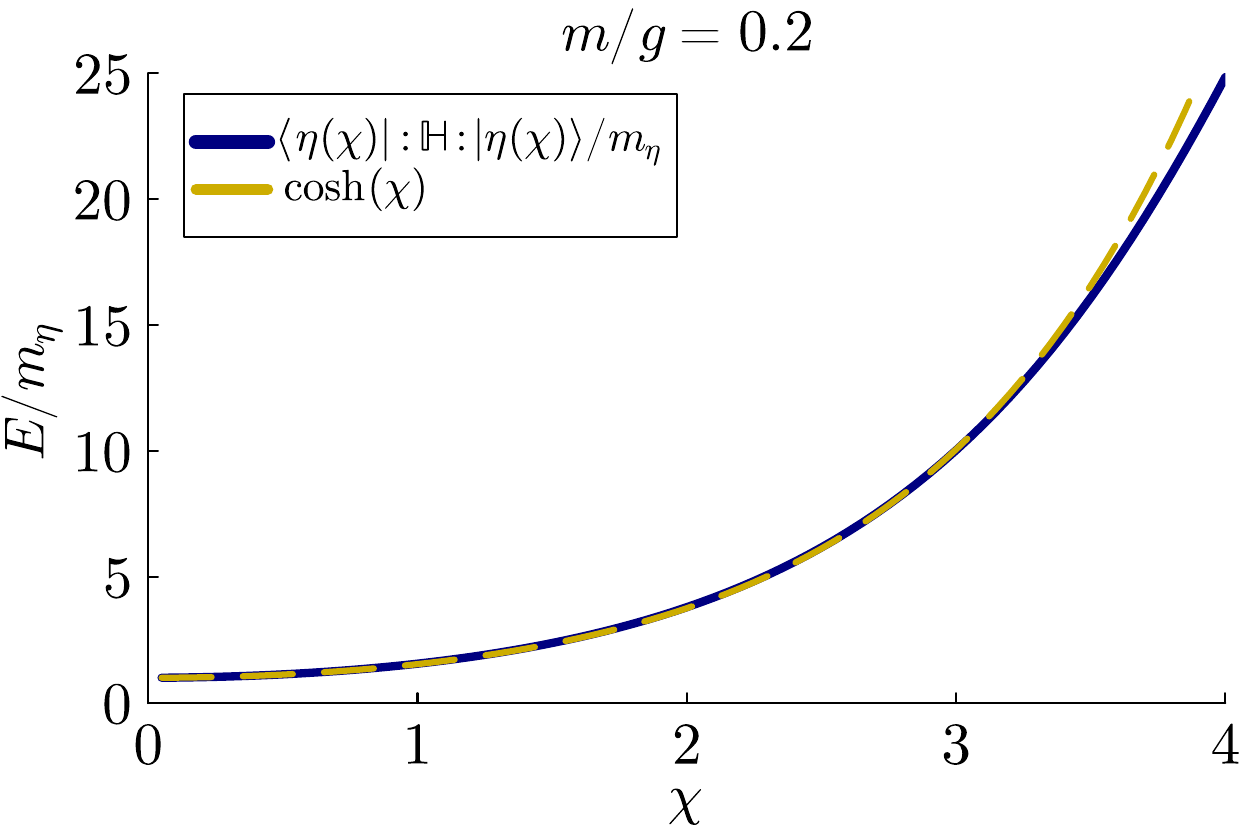}
    \includegraphics[width=.475\linewidth]{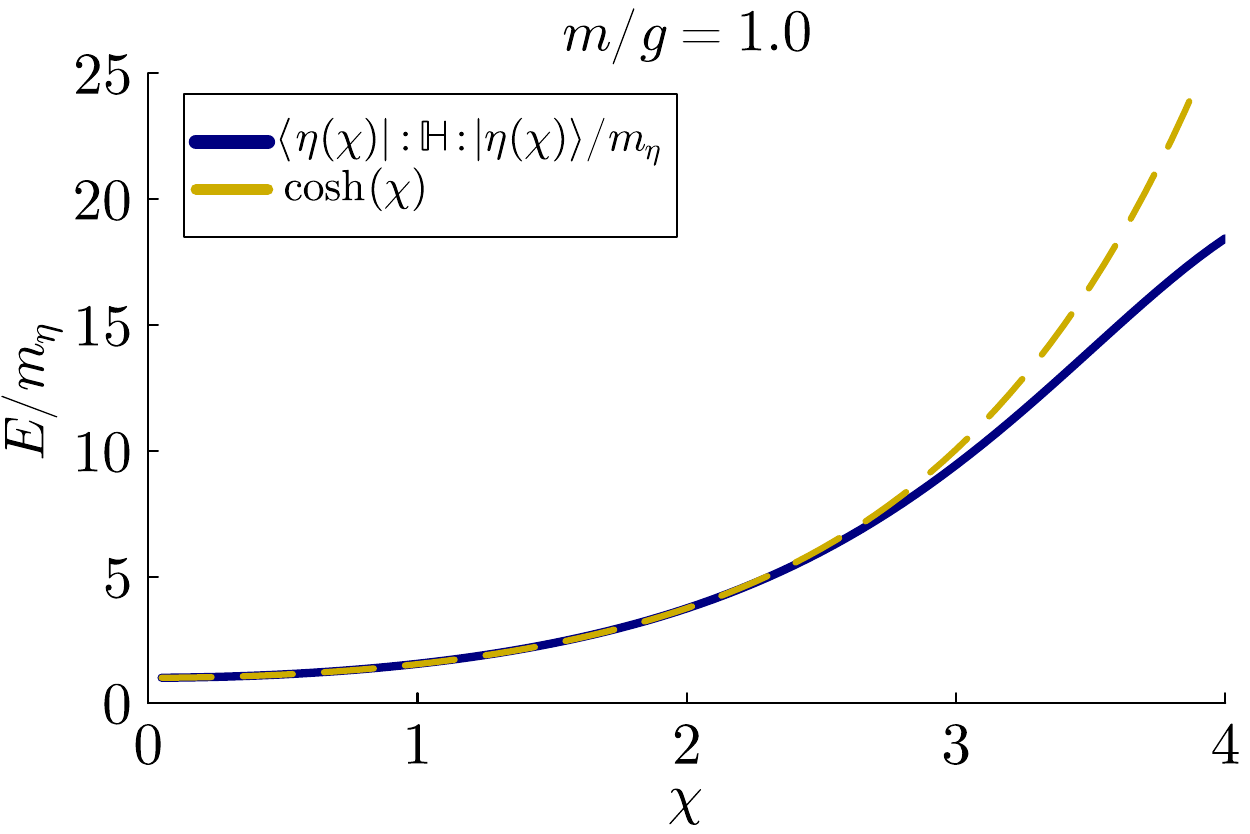}
    \\
    \includegraphics[width=.475\linewidth]{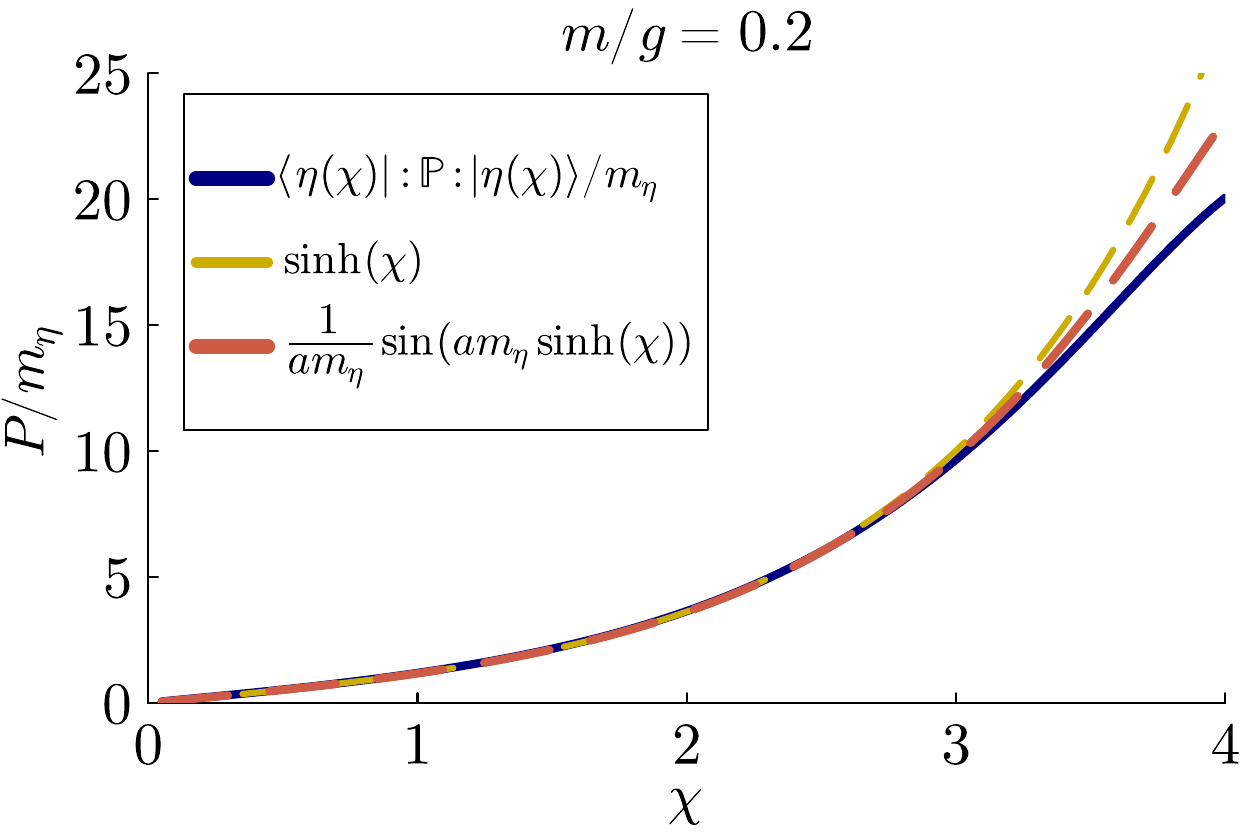}
    \includegraphics[width=.475\linewidth]{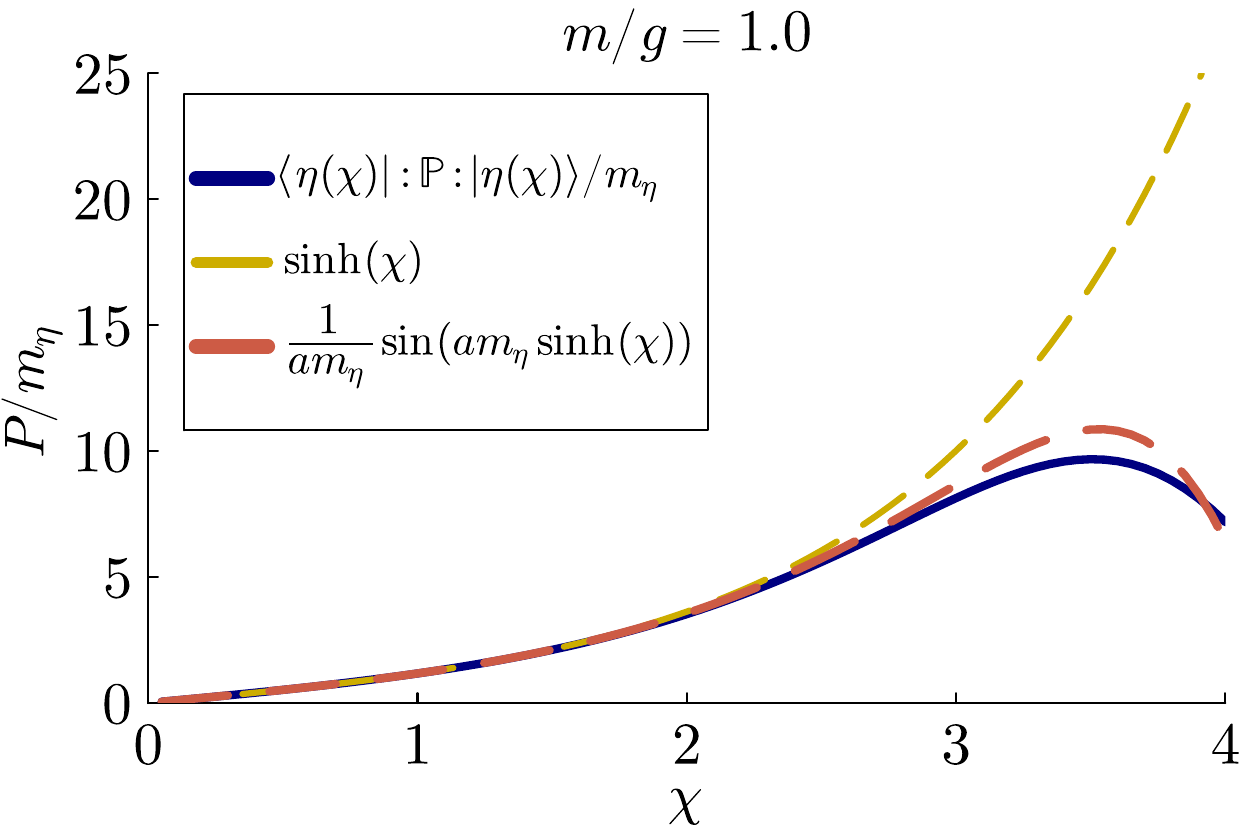}
    \captionsetup{justification=justified}
    \caption{The (unitless) energy and momentum dispersion relations as a function of rapidity $\chi$ with a Trotter step $\delta \chi = 0.05$, where $\ket {\eta(\chi)} = e^{i\mathbb K \chi}\ket\eta$.         The top and bottom rows are the energy and momentum dispersions, respectively, while the left and right columns correspond to the dispersions for $m/g = 0.2$ and $m/g = 1.0$, respectively. The lattice energy expectation value is compared to the continuum result. The lattice momentum expectation value is compared with the continuum result as well as the lattice dispersion relation for an infinite spatial lattice, $\frac 1a \sin(a P(\chi))$.}
    \label{fig:EP_dispersions}
\end{figure*}

The calculations presented in the following sections require the preparation of the vacuum state $\ket\Omega$ and the excited state $\ket\eta$, which are prepared in good approximation with quantum-number conserving MPS and the DMRG algorithm. The vacuum state is prepared by minimizing the expectation value of the lattice Hamiltonian (\ref{eq:HamX}) within the charge zero sector of the Hilbert space. The $\eta$ state, the first charge-zero excited state, is prepared similarly, but a term $\lambda | \!\bra \Omega  \psi\rangle|^2$ is added to the minimized target function, for a heuristically chosen weight $\lambda$. A weight of $\lambda = 10^9$ is used to ensure a small overlap between the prepared states $ |\langle \Omega|\eta\rangle|\sim 10^{-15}$. The DMRG algorithm is used for $\sim 50$ sweeps for sequential singular value cutoffs of $10^{-8}, 10^{-10}, 10^{-14}$, which converges to an MPS with a maximal bond dimension $\sim 500$. Note that this bond dimension is not uniform across the lattice, and the maximal bond is reached only near the center of the system. The mass gap, $m_\eta$, for the light and heavy mass systems is $0.4747$ and $1.2265$, respectively. In the former case $\delta m^2 = m_\eta^2 - (2m_S)^2 >  (2 m_S)^2$, indicating that this system is in the strongly coupled nonperturbative regime. Instead, the heavy-mass system is in the weakly coupled regime. The variances of the lattice Hamiltonian (\ref{eq:HamX}), $\delta E_\psi = (\langle\psi|\mathbb H ^2|\psi \rangle - \langle\psi|\mathbb H|\psi \rangle^2 )^{1/2}$, for both states and for each system's mass parameter are listed in Table~\ref{tab:variances}, and serve as a measure of how close the prepared states are to true eigenstates of the Hamiltonian. The variances are orders of magnitude smaller than the mass gap and, therefore, the prepared states are very tightly centered around the exact eigenstates, with very little contribution from other charge-zero states in the Hilbert space. We note that, in practice, the smallest variances achievable will scale as $O(N)$. This is because the energy expectation will also scale as $O(N)$ and $\langle H^2\rangle \sim O(N^2)$. The difference between two large quantities can only be as small as the floating-point precision. Therefore, floating point errors will scale as $O(N)$ and will limit the lower bound of the achievable variance. 

\subsection{Relativistic dispersion relations}\label{SEC6:Rel_disp}

In (1+1)-dimensional continuum field theory, the energy and momentum operators are obtained from the stress–energy tensor via
\begin{equation}
\int dx^1 T^{0 \mu}(0,x^1) = (\mathbb H,\mathbb P)\,,    
\end{equation}
and transform covariantly under Lorentz boosts generated by $\mathbb U(\chi) \equiv \exp(-i \chi\mathbb K)$.
For a boosted $\eta$ state, the normal-ordered expectation values satisfy 
\be
\begin{split} \label{eq:cont_disp}
    & \langle \eta| \mathbb U(\chi)\! :\!\mathbb H\!: \! \mathbb U^\dagger(\chi)|\eta\rangle = m_\eta \cosh(\chi)\,, \\
    & \langle \eta| \mathbb U(\chi)\! :\!\mathbb P: \! \mathbb U^\dagger(\chi)|\eta\rangle = m_\eta \sinh(\chi) \, ,    
\end{split}
\ee
which implies the standard dispersion relation $\langle : \mathbb  H:\rangle^2 = \langle: \mathbb  P:\rangle^2 +m_\eta^2$. On the lattice, Lorentz symmetry is broken, and this relation is recovered only approximately in the continuum and infinite-volume limits. 

\begin{figure*}
    \centering
    \includegraphics[width=0.475\linewidth]{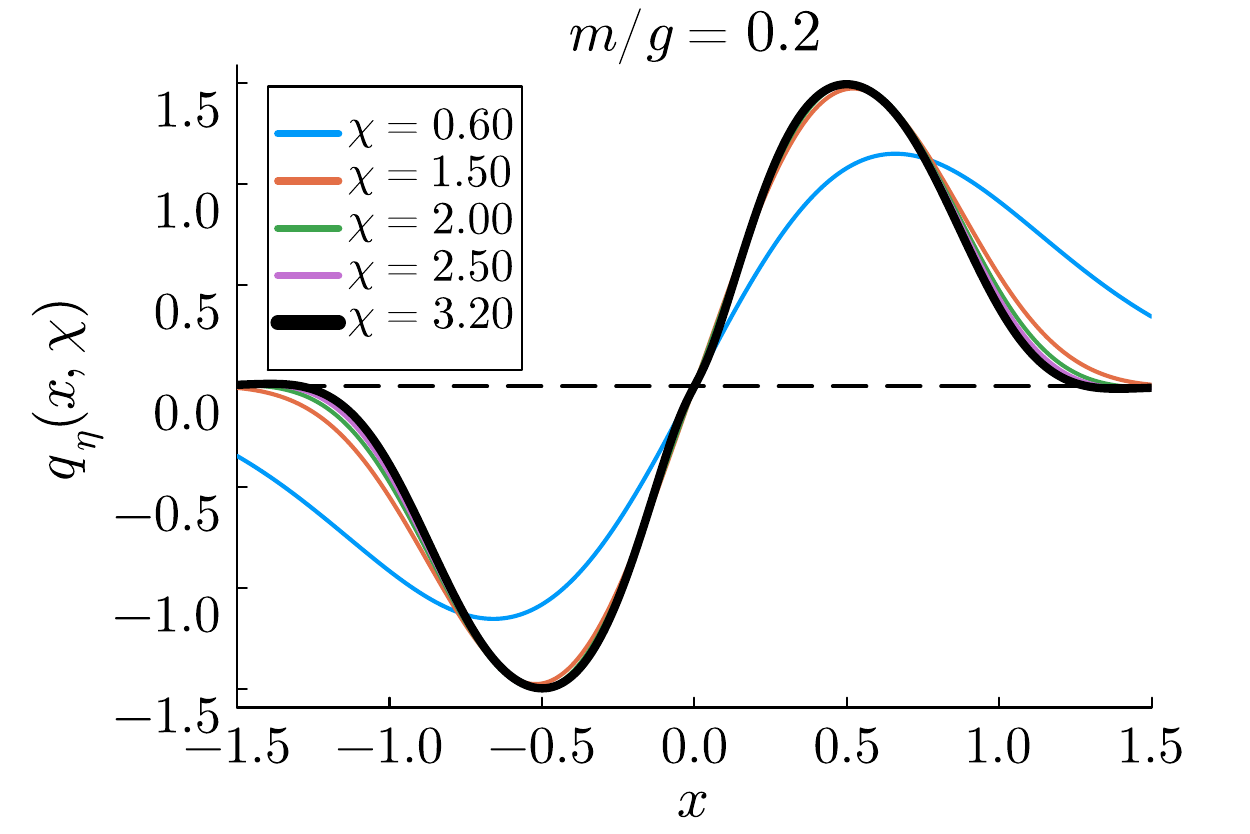} 
    \includegraphics[width=0.475\linewidth]{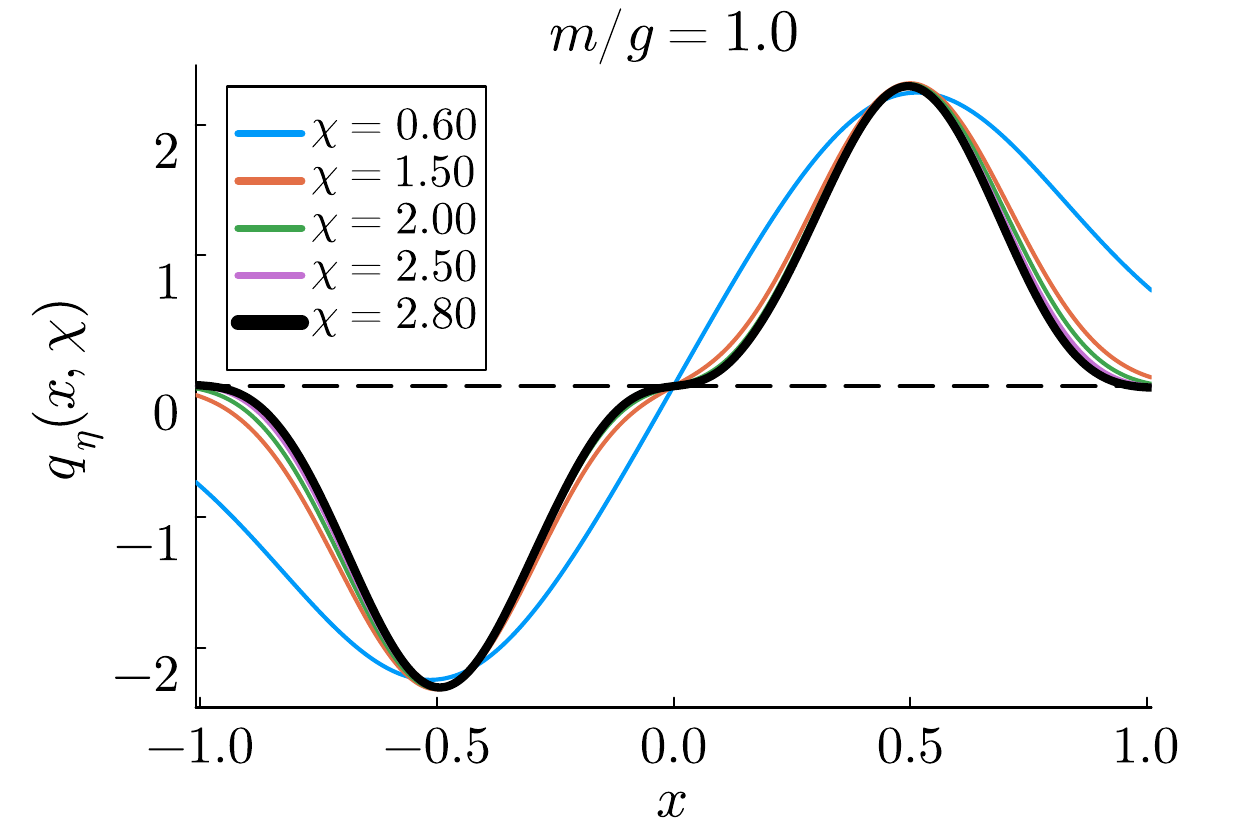}\\
    \includegraphics[width=0.475\linewidth]{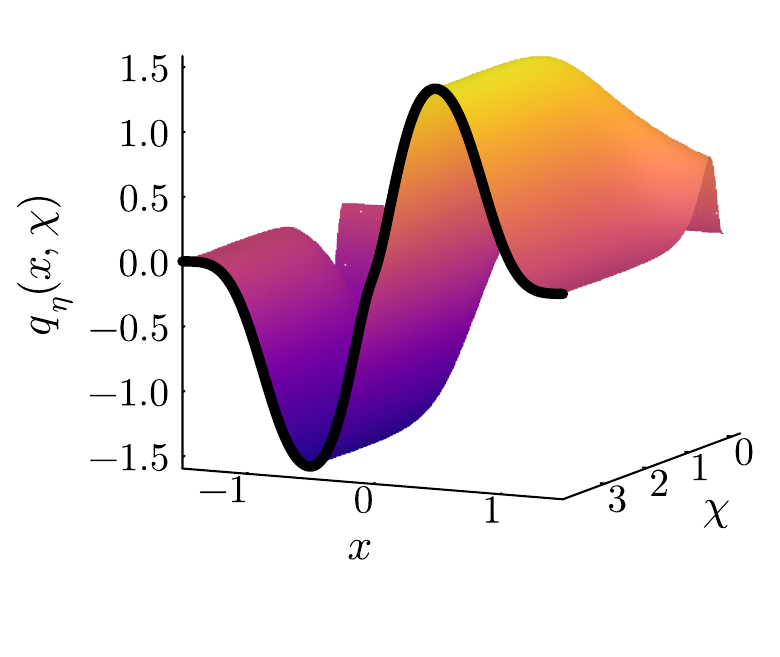} 
    \includegraphics[width=0.475\linewidth]{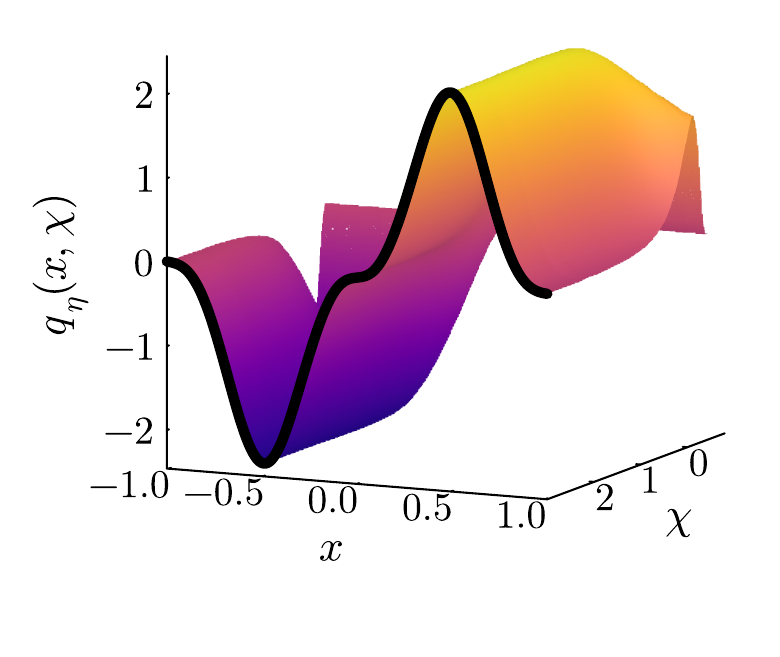}
    \captionsetup{justification=justified}
    \caption{Tensor network calculation of qPDFs for a light mass system (left column) and a heavy mass system (right column). {\bf Top row:} Convergence of the qPDF to the lightcone for different values of the boost rapidity. {\bf Bottom row:} 3D surface plots of the qPDFs in terms of the momentum fraction $x$ and the boost rapidity $\chi$.}
    \label{fig:qPDFs}
\end{figure*}

To validate the boosting procedure used in the qGPD and qPDF calculations, we benchmark the lattice results against the continuum dispersion in Eq.~(\ref{eq:cont_disp}). 
The MPS states $\ket\Omega$ and $\ket{\eta}$ are unitarily evolved using the TDVP algorithm for several rapidities with a second-order Trotter scheme using a cutoff for singular values of $5 \cdot 10^{-8}$. These dispersion relations are presented in Fig.~\ref{fig:EP_dispersions}. While both curves agree well up to some finite value of $\chi$, we generally find a somewhat worse agreement for the momentum than the energy of the boosted state. This can be attributed to the initial states being prepared to be close to eigenstates of $\mathbb H$ while there does not exist a simultaneous diagonal basis for $\mathbb H$ and $\mathbb P$. Therefore, initially, the momentum already has some variance. Since the momentum dispersion performs worse, we use it to determine the maximal value of $\chi$ that is considered in the calculation of the qPDF and qGPD. The larger mass system strays from the continuum dispersion at lower $\chi$. For the qPDF, we choose a maximum rapidity of $\chi = 3.2$ for the lighter mass and $\chi = 2.8$ for the larger mass systems such that the fractional error is $\lesssim 0.05$. These cutoff rapidities correspond to $v \approx 0.9967$ and $v \approx 0.9926$ for the light and heavy mass systems, respectively. 

\subsection{Quasi-PDFs}

The quasi-PDF arises as the $\xi=0$ limit of the quasi-GPD in Eq.~(\ref{eq:qGPD_staggered}). Figure~\ref{fig:qPDFs} shows the resulting quasi-PDFs as functions of the momentum fraction $x$ and the boost rapidity $\chi$. The curves are normalized such that $\int_0^1 dx \;q(x,\chi_\text{max}) = 1$. A detailed discussion of normalization from the tensor-network calculation is given below. Although the lattice dispersion relation breaks down at sufficiently large $\chi$, the quasi-PDFs converge well before this point. To quantify the rapidity convergence, we evaluate the average squared norm of the derivative with respect to $\chi$ at the cutoff rapidities $\chi_{\max}=3.2$ (light mass) and $\chi_{\max}=2.6$ (heavy mass)
\begin{equation}
\begin{split}
        \delta q^2(\chi) &= \frac{1}{2}\int_{-1}^1 dx \bigg| \frac{ \partial q_\eta (x ,\chi) }{\partial \chi} \bigg|^2\\
        &\approx \frac{1}{2}\int_{-1}^1 dx \bigg| \frac{ q_\eta (x ,\chi) -q_\eta(x,\chi- \delta\chi) }{\delta \chi} \bigg|^2 .
\end{split}
\end{equation}
The errors for the PDFs at the cutoff rapidities are $\delta q^2(\chi_\text{max}) = 9.54 \cdot 10^{-4}$ and $\delta q^2(\chi_\text{max}) = 1.32 \cdot 10^{-3}$ for the light and heavy mass systems, respectively, indicating the qPDFs are well converged. 

Both systems display the expected qualitative structure. The qPDF is antisymmetric in $x$ with the particle distribution being the function in the region $x>0$ while the antiparticle distribution is the function in the region $x<0$, with an additional factor of $-1$. For both systems, the distributions are peaked at $|x| = 0.5$, and the support is mostly in the physical region $\abs x \leq 1$, with small lattice-induced tails outside. The heavy-mass system is more tightly confined within the physical domain.

\begin{figure*}
    \includegraphics[width=.475\linewidth]{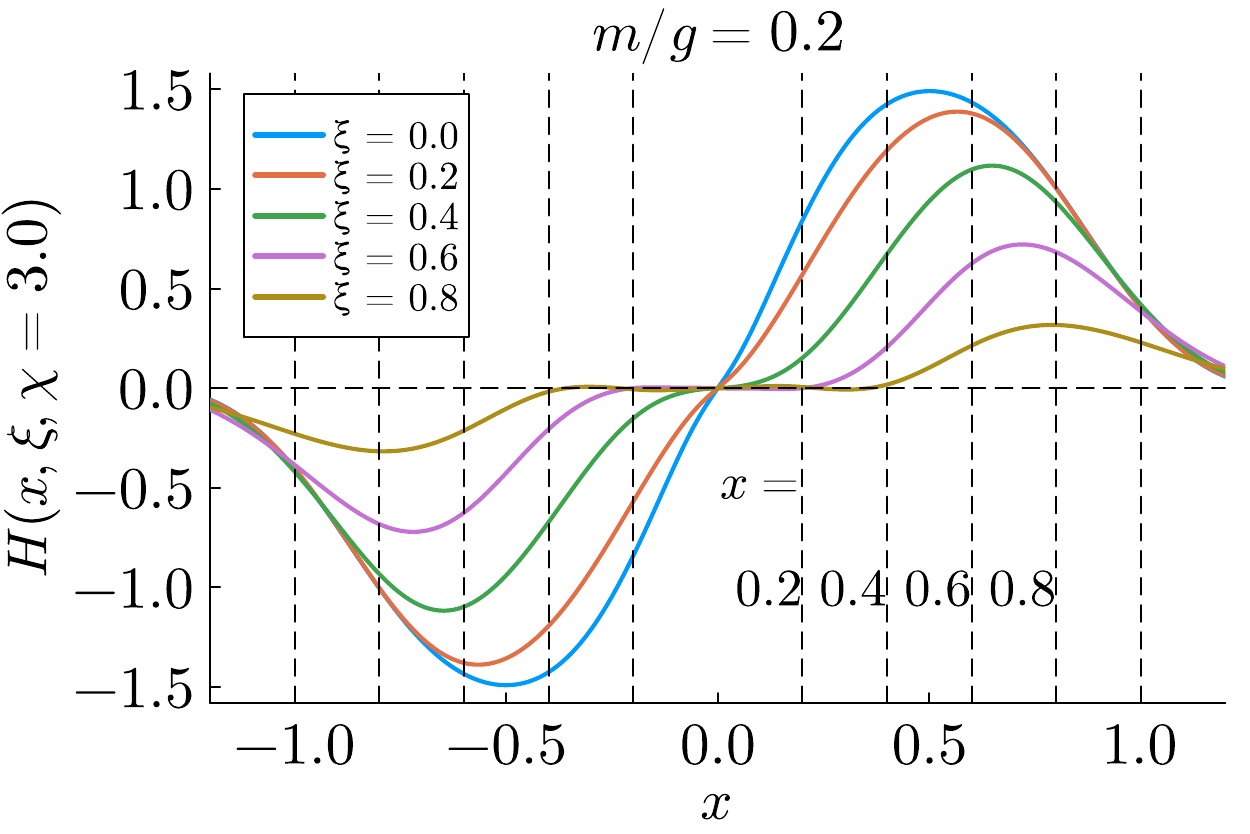} 
    \includegraphics[width=.475\linewidth]
    {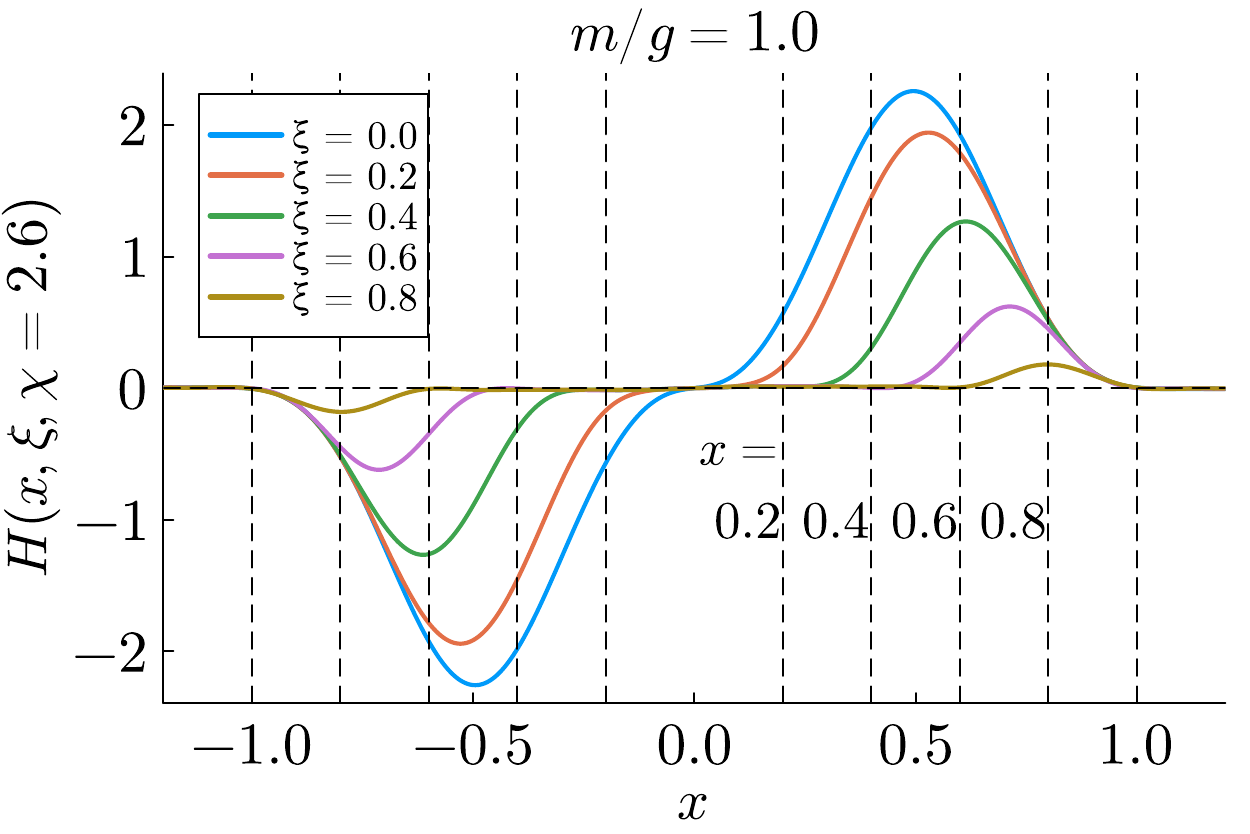}\\
   \hspace{20pt}\includegraphics[width=.475\linewidth]{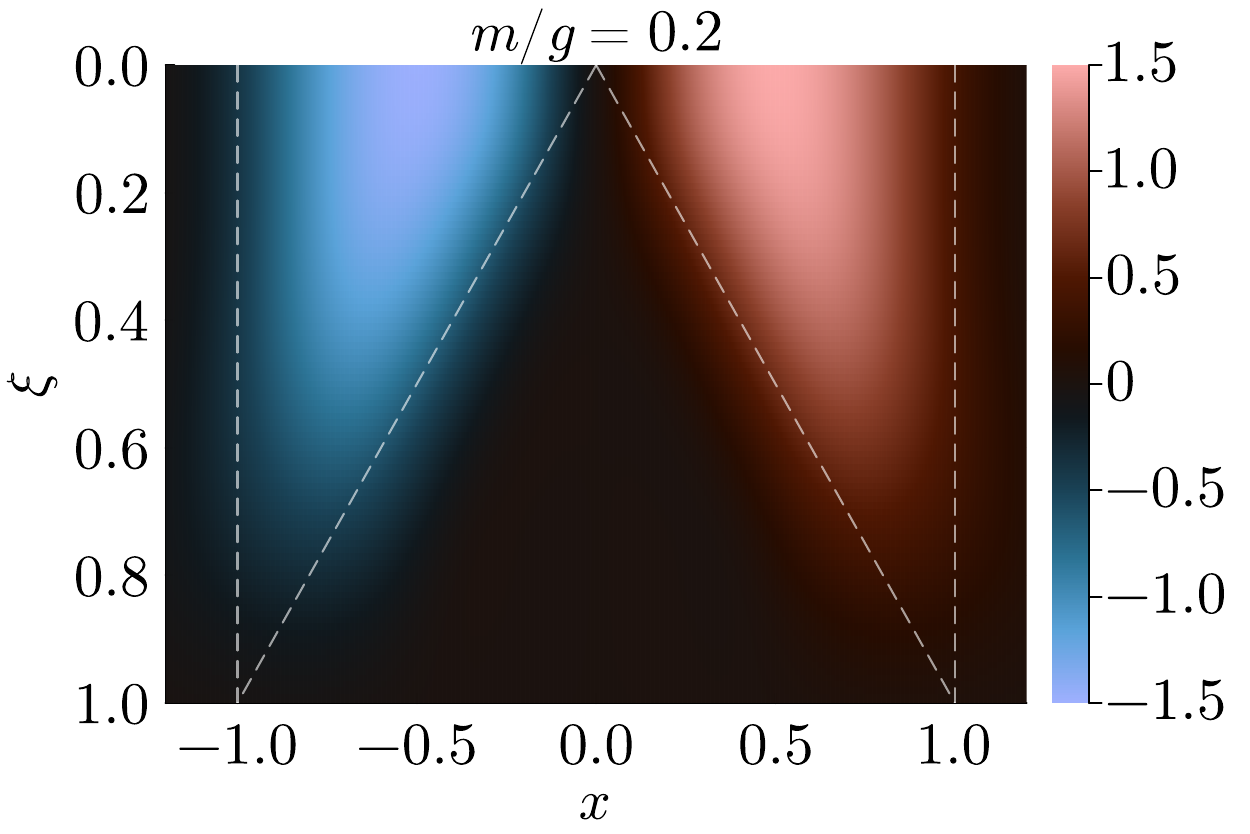}
    \includegraphics[width=.475\linewidth]{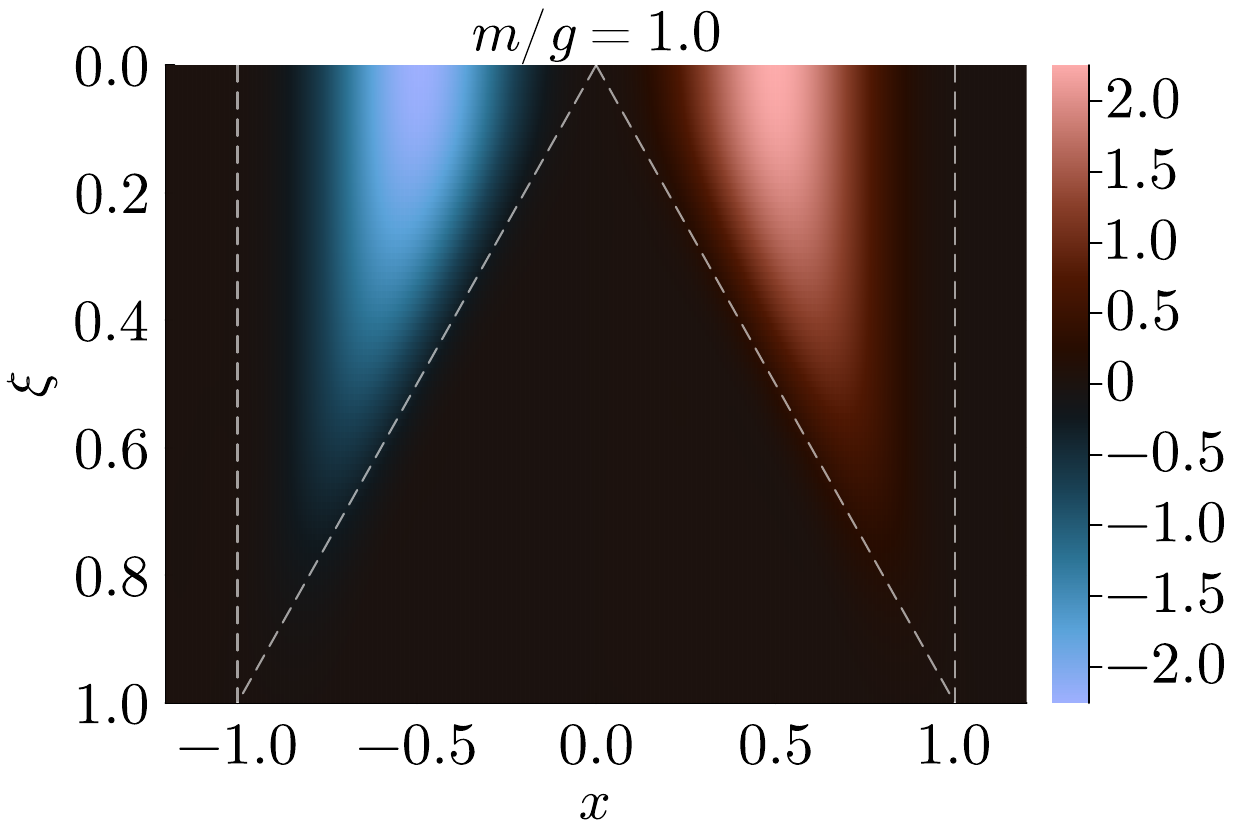}\\
    \includegraphics[width=.475\linewidth]{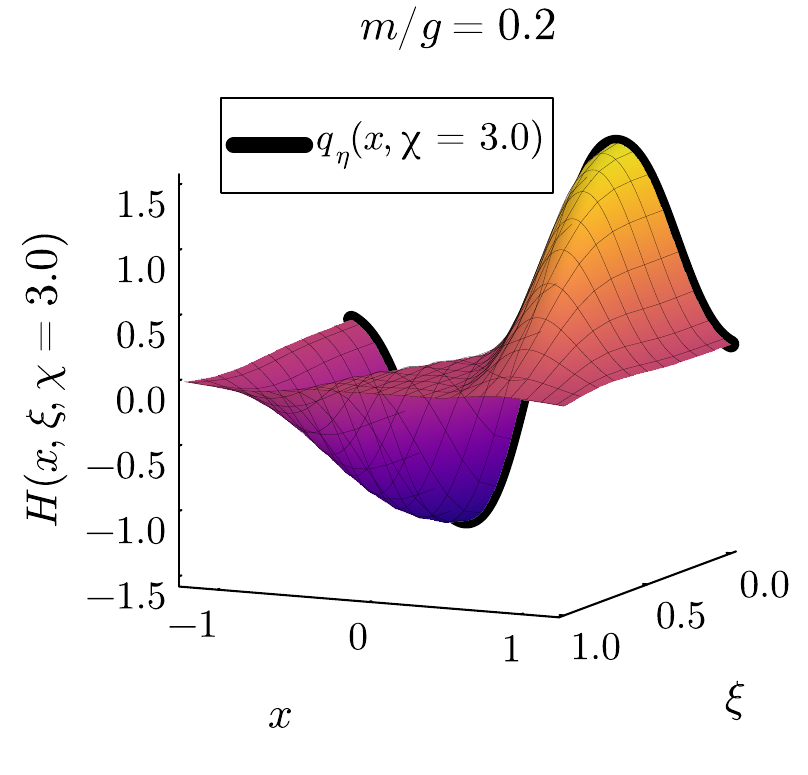}
    \includegraphics[width=.475\linewidth]{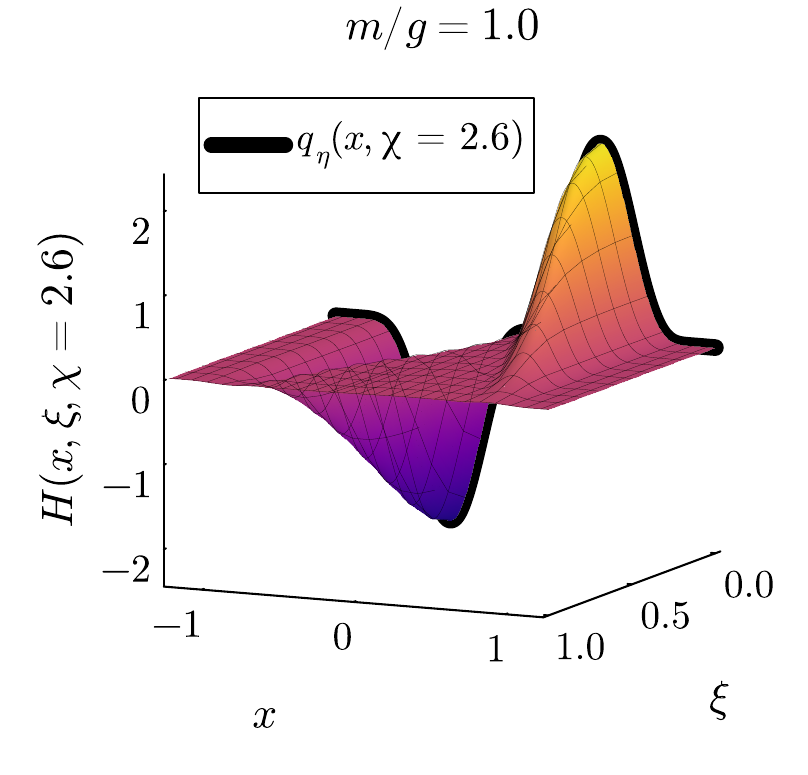}
    \captionsetup{justification = justified, singlelinecheck=false, width = \textwidth}
    \caption{Tensor network calculations of the qGPD for the case of a light mass system (left column) and a heavy mass system (right column). {\bf Top row:} The qGPDs as a function of $x$ for different values of the skewness $\xi$, as indicated in the figure. The dashed vertical lines show the kinematic boundary $|x| = \xi$. 
    {\bf Middle row:} Heatmaps of the qGPD as a function of $x,\xi$, where the dashed line indicates the kinematic boundaries $|x|=\xi$. {\bf Bottom row:} 3D surface plots of the qGPDs. The qPDF, indicated by the black line, is obtained in the limit $\xi =0$.}
    \label{fig:GPD_ft}
\end{figure*}

Light-front PDFs satisfy the following normalization and momentum sum rules~\cite{collins2011foundations}, which we use to quantify the amount of the quasi-distribution that lies outside the region $|x|\leq 1$: 
\begin{eqnarray}
    \sum_{i}\int_0^1dx \,q_i (x) = 2 \label{eq:PDF_norm} \\
    \sum_{i}\int_0^1dx \, x \,q_i (x) =1  \label{eq:PDF_Momumtum_sum}
\end{eqnarray}
These integrals using the distributions (before normalizing) are computed numerically with finite sums taken with enough sample points to converge within the presented precision. We also compute the average momentum fraction $x$ for the distributions $\langle x \rangle = \sum_i\int_x x q_i(x)  / \int_x  q_i(x) $. This average should be exactly $\langle x\rangle =\frac12$ in the infinite volume and continuum limits since there are only two quarks in the system. The results for the integrals and the normalized average $x$ are listed in Table~\ref{tab:PDF_numbers}. The heavy-mass system, which exhibits better localization within the physical kinematic region, satisfies the light-front sum rules more accurately.

\begin{table}[]
    \centering
    \begin{tabular}{c|c|c}
         & $m/g=0.2$ & $m/g=1.0$ \\ \hline
       $\sum_i\int_x  q_i(x) $ (\ref{eq:PDF_norm}) & 4.6785  & 2.0185 \\
        $\sum_i\int_x x q_i(x)$ (\ref{eq:PDF_Momumtum_sum})& 2.4717 & 0.9979 \\
         $\langle x \rangle$ &   0.5273  & 0.4965
    \end{tabular}
    \caption{The qPDF normalizations and momentum sum integrals before normalizing the quasi-PDFs, as well as the average momentum fraction $x$ for the normalized distributions.}
    \label{tab:PDF_numbers}
\end{table}

\subsection{Quasi-GPDs}

The qGPDs for both mass values, computed from Eqs.~(\ref{eq:qGPD_staggered}) and (\ref{eq:JW_op_insertion}), are shown in Fig.~\ref{fig:GPD_ft} for fixed rapidity~$\chi$. For the light and heavy mass systems, we use $\chi = 3.0$ and $\chi = 2.6$, respectively, ensuring that the calculation does not rely on boosted states that deviate significantly from the continuum dispersion relation. In the continuum, we expect GPDs to be depleted outside the DGLAP region, $x \in [-1,-\xi] \cup [\xi,1]$. On the lattice, this support is not imposed by construction. Nevertheless, the numerically obtained qGPDs are predominantly confined to this physical region, with only small residual tails extending into the ERBL regime and beyond $|x|>1$. The lattice results also reproduce the expected qualitative behavior: as $\xi \to 1$, the distributions broaden and decrease in magnitude, consistent with the structure of light-front GPDs. We additionally observe a flattening of the qGPDs within the ERBL region. This behavior is qualitatively compatible with the Reggeized picture discussed in Fig.~\ref{fig:wkb}, although a sharp cusp is not visible at our current lattice spacing and rapidities. Systematic improvements, such as reducing the lattice spacing, increasing the system size, and accessing larger boosts, should lead to closer agreement with the continuum light-front GPDs and sharpen these features.

\section{Conclusions}
\label{SEC7}

In this work, we carried out the first nonperturbative computation of quasi-Generalized Parton Distributions (quasi-GPDs) in the massive Schwinger model, using tensor networks within the Hamiltonian formulation of lattice field theory.  By preparing the first excited mesonic state and boosting it to finite momentum, we were able to evaluate spatial correlation functions that interpolate continuously to their light-front counterparts in the luminal limit.  This approach enables a controlled study of partonic observables in a fully nonperturbative setting, providing a valuable benchmark for the development of ab initio methods aimed at extracting parton physics from first principles. 

The Matrix Product State (MPS) formalism proved suitable for constructing and evolving low-lying states of QED$_2$ in the strongly coupled regime.  We showed that good results can be obtained on lattices of up to 400 sites and that the extracted quasi-distributions exhibit stable convergence with increasing momentum boost. Our numerical findings confirm the consistency of the boosted dispersion relation and provide quantitative evidence that quasi-GPDs reproduce light-front GPDs in the appropriate limit. In addition, analytic results derived from the two-body Fock-space approximation and the Reggeized WKB analysis offered qualitative insight into the observed behavior, especially the depletion of the ERBL region and the emergence of a cusp structure at $x=\pm\xi$.

The combination of tensor-network simulation and analytic modeling thus yields a coherent picture of generalized parton dynamics in two-dimensional QED.  The fair agreement between numerical and analytical trends validates the use of low-dimensional gauge theories as testing grounds for nonperturbative methods that are pertinent to QCD.  Our analysis establishes a set of computational benchmarks that can guide both classical and quantum simulations targeting hadronic observables, including quasi-PDFs and quasi-GPDs.

Future extensions of this work will focus on several directions. On the theoretical side, including higher Fock components and refining the continuum extrapolation will allow for improved resolution of small-$x$ and large-skewness behavior.  From a computational standpoint, the tensor-network framework can be generalized to simulate higher-dimensional and non-Abelian gauge theories, offering a path toward realistic studies of QCD.  Finally, implementing these methods on near-term quantum devices could provide an avenue for demonstrating quantum utility in the calculation of real-time, nonperturbative observables in gauge field theories.

Overall, the present study illustrates that tensor-network simulations of QED$_2$ can capture the essential features of generalized parton structure and offer a tractable, systematically improvable framework for exploring hadronic observables beyond perturbation theory.

\begin{acknowledgments}\noindent
We would like to thank David Kaplan, Nobuo Sato and Zhite Yu for helpful discussions. 
This work is supported by the Office of Science, Office of Nuclear Physics, U.S. Department of Energy under Grants No. DE-FG02-97ER-41014 (UW Nuclear Theory, S.G.), No. DE-FG-88ER40388 (SBU Nuclear Theory) and DE-SC0025881. This work is also supported by the U.S. Department of Energy, Office of Science, National Quantum Information Science Research Centers, Co-design Center for Quantum Advantage (C2QA) under Contract No. DE-SC0012704 (S.G.). This work was supported in part by the U.S. Department of Energy, Office of Science, Office of Nuclear Physics, Inqubator for Quantum Simulation (IQuS) under Award Number DOE (NP) Award DE-SC0020970 (S.G.). This work is in part supported by the Quark-Gluon Tomography (QGT) Topical Collaboration, with Award DE-SC0023646. S.G. was supported in part by a Feodor Lynen Research fellowship of the Alexander von Humboldt Foundation.  This research used resources of the National Energy Research Scientific Computing Center (NERSC), a DOE Office of Science User Facility using NERSC award NP-ERCAP0033891. The authors would like to thank Stony Brook Research Computing and Cyberinfrastructure, and the Institute for Advanced Computational Science at Stony Brook University for access to the SeaWulf computing system, made possible by grants from the National Science Foundation ($\#$1531492 and Major Research Instrumentation award $\#$2215987), with matching funds from Empire State Development’s Division of Science, Technology and Innovation (NYSTAR) program (contract C210148). 
\end{acknowledgments}

\appendix

\section{ Harmonic approximation of the PP-Kernel}
\label{app:harmonic}
Consider the confining PP-kernel
\begin{equation}
\mathcal{K}[\phi](\zeta)
={\rm PP}\!\int_{-1}^{1}\!d\zeta'\,
\frac{\phi(\zeta')-\phi(\zeta)}{(\zeta'-\zeta)^2},\qquad \zeta\in(-1,1).
\end{equation}
For slowly varying $\phi$, we can use the gradient expansion in the kernel 
\begin{align}
\phi(\zeta')=&\,\phi(\zeta)+(\zeta'-\zeta)\phi'(\zeta)\nonumber\\
&\,+\tfrac12(\zeta'-\zeta)^2\phi''(\zeta)+\cdots
\end{align}
to develop a harmonic approximation valid for our WKB estimate. The odd terms vanish under the PP integral, leaving
\be
\mathcal{K}[\phi](\zeta)\approx
\frac12\phi''(\zeta)\!\int_{-1}^{1}\!d\zeta'=\phi''(\zeta),
\ee
so the local limit is
\begin{equation}
\mathcal{K}[\phi](\zeta)\simeq \partial_\zeta^2\phi(\zeta).
\label{eq:local-grad}
\end{equation}
To estimate the harmonic correction on the finite interval, we use the Chebyshev expansion 
\begin{equation}    
\phi(\zeta)=\sum_{n\ge0}a_n T_n(\zeta) 
\end{equation}
with
$T_n(\zeta)=\cos(n\arccos\zeta)$.  The kernel acts diagonally,
\be
{\rm PP}\!\int_{-1}^1\!\frac{T_n(\zeta')-T_n(\zeta)}{(\zeta'-\zeta)^2}d\zeta'
=\tfrac{\pi^2}{2}n^2T_n(\zeta)+\cdots.\hspace{10pt}
\ee
To regulate the divergent $\sum n^2$ sums, we introduce the Abel regulator
\begin{equation}
\Phi(r,\theta)=\sum_{n\ge0}a_n(-1)^{\lfloor n/2\rfloor}r^n\cos(n\theta),\quad 0<r<1,
\end{equation}
and represent the PP-kernel as
\be
\mathcal K_r(\theta)=\frac{\pi^2}{2}(r\partial_r)^2\Phi(r,\theta)+\mathcal R_r(\theta),
\ee
where $\mathcal R_r$ contains the non-diagonal remainder.  The physical kernel
is obtained by taking the Abel limit $r\!\to\!1^-$ after cancellation of divergent pieces.

To evaluate the curvature correction, we carry out a small-angle expansion using $\theta=\tfrac{\pi}{2}+\delta$ ($|\delta|\ll1$), and use
\be
\cos(2m\theta)=(-1)^m(1-2m^2\delta^2+\cdots),
\ee
with $\cos((2m{+}1)\theta)={\cal O}(\delta)$, so that
\begin{align}
[\mathcal D_r]_{\delta^0}
&=\frac{\pi^2}{2}\sum_{m\ge0}(2m)^2a_{2m}(-1)^m r^{2m},\\
[\mathcal D_r]_{\delta^2}
&=-\frac{\pi^2}{2}\delta^2\sum_{m\ge1}(2m)^4a_{2m}(-1)^m r^{2m}.
\end{align}
The remainder $\mathcal R_r$ produces a $+\frac{\pi^2}{2}\delta^2\sum (2m)^4a_{2m}$ term that
cancels the divergent diagonal piece and leaves only $n^2$–weighted sums
\begin{align}
\label{eq:delta2-net}
[\mathcal D_r+\mathcal R_r]_{\delta^2}
=&\,-\pi^2\delta^2
\!\bigg[\sum_{m\ge1}(2m)^2a_{2m}(-1)^m r^{2m}\nonumber\\
&+\sum_{m\ge0}(2m{+}1)^2a_{2m+1}(-1)^m r^{2m+1}\bigg].
\end{align}
The Abel-regularized sums in \eqref{eq:delta2-net} are
\begin{align}
S_{\text{even}}(r)=&\, \sum_{m\ge1}(2m)^2(-1)^m r^{2m}\,,\nonumber\\
S_{\text{odd}}(r)=& \,\sum_{m\ge0}(2m{+}1)^2(-1)^m r^{2m+1}\,.
\end{align}
Although $S_{\text{even,odd}}(r)\to0$ as $r\!\to\!1^-$,
their \emph{Abel integrals} are finite:
\begin{align}
\int_0^1\!\frac{dr}{r}S_{\text{even}}(r)&=-\frac{\pi^2}{24},&
\int_0^1\!\frac{dr}{r}S_{\text{odd}}(r)&=-\frac{\pi^2}{8}.
\end{align}
The sum of these contributions gives the universal constant
\be
-\frac{\pi^2}{24}-\frac{\pi^2}{8}=-\frac{\pi^2}{6}.
\ee
This constant multiplies the even projection of $\phi$ at $\zeta=0$, since
\begin{equation}
\phi(0)=\sum_{m\ge0}a_{2m}(-1)^m,\qquad
\sum_{m\ge0}a_{2m+1}(-1)^m=0,
\end{equation}
reflecting $T_{2m}(0)=(-1)^m$ and $T_{2m+1}(0)=0$, hence
\begin{equation}
\lim_{r\to1^-}[\mathcal D_r+\mathcal R_r]_{\delta^2}
=\frac{\pi^2}{6}\,\delta^2\,\phi(0).
\end{equation}
The use of the Abel sums without the explicit coefficients $a_{2m}$ is justified for smooth $\phi(\zeta)$ in the harmonic approximation or small $\zeta$, since the Chebyshev coefficients $a_n$ decrease rapidly with $n$. In the narrow band of large-$n$ modes dominating the PP-kernel, they can be approximated by their slowly varying average $a_0\simeq\phi(0)$. The Abel regulator isolates the universal, UV–dominated part of the spectral sum, which depends only on the kernel structure (the $\sum (-1)^n n^2r^n$ factors), but not on the detailed envelope of $a_n$. Hence the  average
\be
\sum_{n} a_n(-1)^n n^2 r^n
\simeq a_0\,\sum_n (-1)^n n^2 r^n
\ee
is justified as $a_n$ varies slowly compared with the oscillatory kernel. In a way, this is analogous to extracting a local counter-term coefficient in field theory, where the divergent factor is universal and multiplies the smooth field $\phi(0)$.

With this in mind, using $\delta^2=\zeta^2+\mathcal O(\zeta^4)$ and restoring the gradient term~\eqref{eq:local-grad},
the PP-kernel near $\zeta=0$ takes the local form
\begin{equation}
\mathcal K[\phi](\zeta)
=\partial_\zeta^2\phi(\zeta)
+\frac{\pi^2}{6}\,\zeta^2\,\phi(\zeta)
+\mathcal O(\zeta^3,\zeta^2\phi')\,,
\label{eq:K-local-final}
\end{equation}
after the substitution $\phi(0)\rightarrow\phi(\zeta)$ modulo higher gradients. Eq.~\eqref{eq:K-local-final} shows that the nonlocal confining interaction acts as a harmonic potential $\tfrac{\pi^2}{6}\zeta^2$ near the origin, with the universal curvature $\pi^2/6$ originating from the Abel-summed even/odd spectral constants $\zeta(2)/4$ and $3\zeta(2)/4$.

\section{Discrete symmetries of 2D GPDs}\label{app:discsym}

For the continuum Dirac fields and their currents in 2D, the discrete {\cal C, P, T} transformations 
can be summarized as follows
\begin{itemize}
\item $\cal C$: $\psi\mapsto C\bar\psi^T$, ${C}=\gamma^1$.
     $j^\mu\mapsto -j^\mu$, $j^\mu_5\mapsto +j^\mu_5$.
\item $\cal P$: $x\to -x$, $\psi\to\gamma^0\psi$,,
     $j^0\mapsto j^0$, $j^1\mapsto -j^1$\,,
\item $\cal T$: $t\to -t$, $\psi\to \gamma^0\psi^*$.
\end{itemize}
Note that under charge conjugation $g$ is {\it even}, since the interaction term is invariant
\begin{equation}
-gj^\mu A_\mu\xrightarrow{C} -(+g)(-j^\mu)(-A_{\mu})=-gj^\mu A_\mu.
\end{equation}
One can show that ${\mathbb H,\mathbb K},\mathcal O_\Gamma$ are ${\cal T}$-even for $\Gamma={\bf 1}, \gamma^0,\gamma^5,\gamma^0+v\gamma^1$.
With this in mind, we now discuss the general properties of GPDs under hermiticity and time-reversal-parity ${\cal T_P}={\cal TP}$ for a spinless meson state in 2D. Recall that in 2D ${\cal T}=\gamma^0K$ (anti-unitary)
and ${\cal P}=\gamma^0$ (unitary). We note that the arguments in the following apply both to 2D GPDs on the light cone and qGPDs involving a spatial gauge link.
\\
\\
{\bf 1. Hermiticity:}
\\

The conjugation of the $z$-dependent kernel in the qGPD gives
\begin{align}
H_\Gamma^*(z^-,\xi,t)=&\,\langle p_2|\overline \psi (-z^-)\Gamma\psi(z^-)|p_1\rangle^*\nonumber\\
=&\,\langle p_1|\overline \psi (z^-)\Gamma^\dagger\psi(-z^-)|p_2\rangle\nonumber \\
=&\,H_\Gamma(-z^-,-\xi,t)\,,
\end{align}
or after the Fourier transform
\bea
H_\Gamma^*(x,\xi,t)=H_\Gamma(x,-\xi,t)\,.
\eea
\\
\\
{\bf 2. Time-reversal-parity:}
\\

Time-reversal-parity symmetry applied to the $z$-kernel of the qGPD, gives
\begin{align}
H_\Gamma(z^-,\xi,t)=&\,\langle p_2|{\cal T_P}^\dagger{\cal T_P}\overline \psi (-z^-)\Gamma\psi(z^-)|p_1\rangle^*\nonumber\\
=&\,\langle p_1|\overline \psi (-z^-)\Gamma^\dagger\psi(z^-)|p_2\rangle\nonumber\\
=&\,H_\Gamma(z^-,-\xi,t)\,.
\end{align}
or after the Fourier transform
\bea
H_\Gamma(x,\xi,t)=H_\Gamma(x,-\xi,t)
\eea
The reality of the qGPD follows from the combined identities.

\section{Quantum algorithm for quasi-PDFs and quasi-GPDs}
\label{app:quantum}

In this section, we describe quantum algorithms to evaluate both quasi-PDFs and quasi-GPDs. As discussed, for example, in Ref.~\cite{Li:2021kcs}, PDFs can be accessed with quantum computers using the Hadamard test, since they reduce to expectation values of Pauli strings evaluated on the boosted state. We denote the boosted excited state as $\ket{\eta(\chi)}=e^{i\chi \mathbb K }\ket{\eta}$. The operators ${\cal O}(z)$ relevant for the quasi-PDF calculations can be decomposed in terms of Pauli strings $P_i \in \{I, X, Y, Z\}^{\otimes n} $ as
\begin{equation}
    {\cal O}(z)=\sum_i a_i P_i\,,
\end{equation}
with coefficients $a_i\in \mathbb{R}$. The real space PDF on the lattice, $q_\eta(z,\eta)$, given by the summand of (\ref{eq:qGPD_staggeredII}) for $\xi =0$, then takes the form
\begin{equation}
    \sum_{ij}a_i \tilde a_j\bra{\eta(\chi)} P_iP_j \ket{\eta(\chi)}\,,
\end{equation}
which can be evaluated term by term using the Hadamard test. This is achieved by introducing an ancillary qubit initialized in the $\ket{0}$ state. We execute the following quantum circuit
\vspace{.4cm}
\begin{eqnarray}
\label{eq:had_test}
&& \hspace*{\fill}
\Qcircuit @C=1em @R=1em {
	\lstick{\ket{0}} & \gate{H}&\ctrl{1}&\ctrl{1} & \gate{H} &\meter \nonumber \\
	\lstick{\ket{\eta(\chi)}} & \qw&\gate{P_{j}} &\gate{P_i}&\qw&\qw \;\;\;\;\;\;\;\;\;
}\,\\ 
\end{eqnarray}
where the lower wire represents the set of qubits prepared in the boosted state. Conditional applications of Pauli strings are followed by the measurement of the ancilla in the computational basis. The measurement statistics yield the real part of the desired correlation function 
\begin{equation}
P(\ket{0})- P(\ket{1}) = \text{Re}\left[ \bra{\eta(\chi)}P_i P_j\ket{\eta(\chi)} \right]\,.
\end{equation}
The left-hand side denotes the difference in probabilities of measuring the ancilla in the state $\ket{0}$ and $\ket{1}$. Analogously, the imaginary part of the correlation function is obtained by including an additional phase $S^\dagger$ gate for the ancilla qubit just after the first Hadamard gate.

The quasi-GPD is an off-diagonal matrix element that can be written in the form 
\begin{eqnarray}
    \bra{\eta(\chi)}U^\dagger(\delta_2)P_i\,P_jU(\delta_1)\ket{\eta(\chi)}
\end{eqnarray}
where $U(\theta)= e^{i \theta \mathbb K }$ denotes the unitary boost operator, and $\delta_{1/2}= \chi_{1/2}(\xi, \chi)-\chi$. These matrix elements can be measured using the following quantum circuit
\vspace*{.3cm}
\begin{eqnarray}
\label{eq:had_test1}
&&\;\;\;\;\;\; 
\Qcircuit @C=1em @R=1em {
	\lstick{\ket{0}} & \gate{H}&\ctrl{1}&\ctrl{1}&\ctrl{1}&\ctrlo{1}& \gate{H}&\meter \nonumber \\
	\lstick{\ket{\eta(\chi)}} &\qw &\gate{U_1} & \gate{P_j} & \gate{P_i} & \gate{U^\dagger_2}&\qw&\qw\;\;\;\;\;\;
} \\ 
\end{eqnarray}
where we use the shorthand notation $U_{1/2}=U(\delta_{1/2})$. Similar to the PDF, the difference in probabilities for the outcomes of the ancilla qubit gives the real part of the quasi-GPD matrix element
\begin{align}
P(\ket{0})- P(\ket{1}) = &\, \nonumber \\
&\hspace*{-2cm} \text{Re}\left[ \bra{\eta(\chi)}U^\dagger(\delta_2)P_i\,P_jU(\delta_1)\ket{\eta(\chi)} \right]
\end{align}
The imaginary part of the quasi-GPD can be obtained by inserting an additional phase gate $S^\dagger$. The implementation of the circuit shown in (\ref{eq:had_test1}) requires the conditional application of the unitaries $U_{1/2}$, rather than only conditional Pauli strings. Compared to the PDF case, this leads to a larger number of single- and two-qubit gates when decomposed into elementary operations. We leave a more quantitative assessment for future work.

\bibliography{main}

\end{document}